\documentclass[twocolumn]{aastex701}

\usepackage{comment}
\usepackage{amsmath}
\usepackage{adjustbox}
\usepackage{graphicx}

\begin{document}

\title{Modelling the Time-variable Broadband Emission and Correlation Study of FSRQ S5\,1044+71}



\author[]{Sajad Ahanger}
\altaffiliation{Department of Physics, University of Kashmir, Srinagar, 19006, India}
\affiliation{Department of Physics, University of Kashmir, Srinagar, 19006, India}
\email[show]{sajadphysics@gmail.com}  

\author[]{Shah Zahir} 
\altaffiliation{Department of Physics, Central University of Kashmir, Ganderbal, 191201, India.}
\affiliation{Department of Physics, Central University of Kashmir, Ganderbal, 191201, India.}
\email[show]{shahzahir4@gmail.com}

\author[]{Sunder Sahayanathan}
\affiliation{Astrophysical Sciences Division, Bhabha Atomic Research Centre, Mumbai, 400085, India.}
\affiliation{Homi Bhabha National Institute, Mumbai, 400094, India.}
\email{}

\author{Naseer Iqbal}
\affiliation{Department of Physics, University of Kashmir, Srinagar, 19006, India.}
\email{}

\author[]{Malik Zahoor}
\affiliation{Department of Physics, National Institute of Technology, Srinagar 190006, India.}
\email{}

\author[]{Aaqib Manzoor}
\affiliation{Indian Institute of Astrophysics, Bangalore 560034, India.}
\email{}



\begin{abstract}

We present a detailed temporal and spectral analysis of the blazar S5\,1044+71 using multi-wavelength data obtained from the \emph{Fermi}-LAT and Swift-XRT/UVOT telescopes. Applying the Bayesian block algorithm to the 3-day binned $\gamma$-ray lightcurve, we identify pronounced variability, including four major outbursts marked by significant flux enhancements. The highest flux recorded was $(1.1 \pm 0.2)\times 10^{-6}\,\text{ph}\,\text{cm}^{-2}\,\text{s}^{-1}$ on 57868.5 MJD. Each outburst comprises multiple components, and lightcurve profile analysis indicates predominantly symmetric temporal structures.
The shortest variability timescale of 4.5 hours constrains the emission region to be located within 0.03 pc of the central engine, likely near the broad-line region (BLR). Additionally, two highest-energy photons were detected with energies of 46.4 GeV (on 57739.6 MJD) and 42.5 GeV (on 59161.9 MJD), observed outside the peak flaring activity.
The fractional variability shows an overall increasing trend from UV/optical to $\gamma$-ray bands, with a noticeable dip in the X-ray range, consistent with the shape of the broadband spectral energy distribution (SED). The flux distributions during flares exhibit log-normal or double log-normal behavior, suggesting multiplicative variability processes and evolving emission zones. Cross-correlation analysis reveals a strong positive correlation between the $\gamma$-ray and X-ray bands, with X-rays lagging by 42.5 days. 
Broadband SED modeling across different flux states supports a one-zone leptonic scenario, with $\gamma$-ray emission produced via external Compton scattering of IR and BLR photons. High-flux states show harder electron spectra, elevated break energies, and reduced magnetic fields—features consistent with efficient particle acceleration and Compton dominance.

\end{abstract}

\keywords{radiation mechanisms: non-thermal - galaxies: active - galaxies: individual: S5\,1044+71 - jets: gamma-rays.}


\section{INTRODUCTION}
Blazars, including BL Lac objects and flat-spectrum radio quasars (FSRQs), belong to a subclass of radio-loud active galactic nuclei (AGNs) with a relativistic jet pointing towards the earth \citep{1995PASP..107..803U}. BL Lac objects are characterized by the weak or absence of optical emission lines, in contrast to flat-spectrum radio quasars (FSRQs), which exhibit such lines. The relativistic effects boost the intrinsic emission from blazars, making them some of the luminous objects in the universe. The non-thermal emission from blazars have been detected across the entire electromagnetic (EM) spectrum, with various ground- and space-based telescopes, spanning energy ranges from low-frequency radio to TeV-energy $\gamma$-rays. Moreover, blazars exhibit unique observational features such as large amplitude variability, high and variable polarization, compact radio emission and high flux variability, with a variability timescale of the order of minutes to years \citep{1995PASP..107..803U,2017NatAs...1E.194P,2018Galax...6...34G}. 

The broadband spectral energy distribution (SED) of blazars feature a distinctive double-peaked structure. The low-energy peak, occurring at optical, ultraviolet (UV) to X-ray energies, is attributed to Doppler-boosted synchrotron emission due to low energy electrons in a magnetized jet \citep{1978bllo.conf..328B,ghisellini1989}. The high-energy peak, which appears at $\gamma$-ray energies, can be interpreted through various models. The most widely accepted explanation involves the inverse Compton (IC) process, in which synchrotron photons, external photons, or both are upscattered to higher energies by relativistic particles within the jet. In case synchrotron photons serve as the seed photons for IC scattering, the process is known as synchrotron self-Compton (SSC) \citep{1985A&A...146..204G,1992ApJ...397L...5M,1996ApJ...461..657B}. If the seed photons originate outside the jet, such as from accretion disk, broad line region (BLR) or dusty torus, the process is referred to as external Compton (EC)  \citep{1993ApJ...416..458D,1994ApJ...421..153S,2017MNRAS.470.3283S,2018MNRAS.477.4749C}. Apart from leptonic models, another possible explanation for the high-energy component involves proton-synchrotron processes or proton cascades \citep{1992A&A...253L..21M,1999MNRAS.302..373B,2001APh....15..121M,2000NewA....5..377A,2020ApJ...889..149D}. These processes are believed to be responsible for producing not only $\gamma$-ray photons but also high-energy neutrinos, therefore making lepto-hadronic models more favorable for explaining the broadband SED of blazars where neutrino detection is reported \citep{2018ApJ...866..109S,2024ApJ...977...42R,2024MNRAS.529.3503B}. Based upon the peak frequency ($\nu_\text{syn}$) of the synchrotron component, blazars are further classified into three types: low synchrotron peaked blazars (LSP; $\nu_\text{syn} \leq 10^{14}$ Hz), intermediate synchrotron peaked blazars (ISP; $10^{14} \leq \nu_\text{syn} \leq 10^{15}$ Hz), and high synchrotron peaked blazars (HSP; $\nu_\text{syn} \geq 10^{15}$ Hz) \citep{2010ApJ...716...30A}.  

Blazar variability has been extensively studied to understand the nature of non-thermal emission and energy dissipation in relativistic jets. The variability timescales in blazars are generally classified into three categories: intraday (hours to days), short-term (days to months), and long-term (months to years). In the $\gamma$-ray band, flare timescales vary from minutes to hours providing insights into the underlying physical properties, such as the location and size of emission region, as well as jet dynamics \citep{2014MNRAS.442..131K,2015ApJ...804...74P,2016ApJ...824L..20A,2019ApJ...881..125D}. However, the exact mechanisms responsible for these extreme $\gamma$-ray flares are still not fully understood. Apart from that, minute-scale variability poses significant challenges in explaining how broadband emission is produced near the supermassive black hole (SMBH) in such a compact region. Additionally, the underlying physical mechanism responsible for accelerating charged particles to relativistic speeds remains a complex issue. A variety of scenarios have been proposed to address these phenomenon in blazars including, shock-in-jet models \citep{2010ApJ...711..445B,2018A&A...619A..45M}, magnetic reconnection events \citep{2013MNRAS.431..355G,2014PhRvL.113o5005G,2019MNRAS.486.1548M}, recollimation shock models \citep{2018A&A...609A.122B}, and cloud-in-jet models \citep{2012ApJ...749..119B,2013MNRAS.436.3626A,2019A&A...623A.101D}. Furthermore, the flaring periods in blazars can vary depending on the underlying mechanism driving the flare. In the leptonic scenario, such flares are generally expected to produce correlated variability across multiple wavebands, often occurring nearly simultaneously. Under this scenario, relativistic electrons in the jet are responsible for both optical and $\gamma$-ray emissions, so a close correlation between their flux variations is expected \citep{2007Ap&SS.309...95B}. Moreover, correlation between different wavebands with or negligible time lag has been found in many blazars \citep{2009ApJ...697L..81B,2012ApJ...749..191C,2014ApJ...783...83L,2018MNRAS.480.5517L}. These results strongly support the one-zone leptonic model of blazar emission. Nevertheless, several studies challenge this model due to the lack of correlation between flux variations among different wavebands. In such studies, there are instances where either $\gamma$-ray flares occur without their corresponding optical/X-ray flares (orphan $\gamma$-ray flares) \citep{2013ApJ...779..174D, 2015ApJ...804..111M, 2020MNRAS.498.5128R}, or optical flares occur without their $\gamma$-ray/X-ray counterparts (orphan optical flares) \citep{2013ApJ...763L..11C, 2019MNRAS.486.1781R, 2020MNRAS.498.5128R}.

S5\,1044+71 is a  distant flat-spectrum radio quasar (FSRQ) at a redshift of z $\sim$ 1.15 \citep{1995ApJS...98....1P}, associated with \emph{Fermi} $\gamma$-ray source 4FGL J1048.4
+7143 \citep{2020ApJS..247...33A,2020arXiv200511208B}. The source  was classified as the low synchrotron-peaked blazar in the second LAT AGN catalog \citep{2011ApJ...743..171A}, with R.A = 162.115083 and Dec. = 71.726649. Large Area Telescope (LAT) recorded  $\gamma$-ray flaring activity from S5\,1044+71 on 2013 March 01, and 2014 January 16-17 \citep{2013ATel.4941....1O,2014ATel.5784....1D}. Furthermore, on 2016 December 29,  the $\gamma$-ray flux of the source increased significantly, reaching a maximum daily averaged flux $(E > 100 \ \text{MeV}) \ \text{of} \ (1.1 \pm 0.2) \times 10^{-6} \ \text{ph} \ \text{cm}^{-2} \ \text{s}^{-1}$, which is about 16 times higher than the average flux reported in the third \emph{Fermi}-LAT catalog (3FGL) \citep{2017ATel.9928....1O}. The source exhibited near-infrared brightening in 2013 January, becoming approximately 1.2 magnitudes brighter than its previous flux \citep{2013ATel.4815....1C}. On 2013 October 25, its R-band flux was observed in a flaring state with about 1.5 magnitudes brighter than its usual level \citep{2013ATel.5512....1B}. Later, it entered a high-radio state between 2014 January and February \citep{2014ATel.5792....1T,2014ATel.5869....1T}. A significant optical enhancement was observed in 2017 January, with an R-band magnitude of $15.44 \pm 0.20$ \citep{2017ATel.9956....1P}, simultaneous with a $\gamma$-ray flare.

FSRQ S5\,1044+71 has been investigated in several studies for possible quasi-periodic oscillations (QPOs). The analysis of a 5-day binned $\gamma$-ray lightcurve for the source revealed a potential QPO of $3.06 \pm0.43 $ years, and the application of a binary black hole model suggests the existence of a binary supermassive black hole at its center \citep{2022ApJ...929..130W}. A study on 24 bright active galactic nuclei by \cite{2023A&A...672A..86R} identified the longest QPO in S5\,1044+71, with a period of approximately 1130 days. \cite{2022ApJ...940..163K} proposed a multi-messenger model for the source, attributing flaring activity to the spin–orbit precession of a binary black hole based on \emph{Fermi}-LAT $\gamma$-ray and VLBI observations. They also expanded on the existing model and predicted the duration time of the next $\gamma$-ray flare. A broadband spectral analysis of the high and low states of 12 distinct samples of \emph{Fermi}-4LAC bright FSRQs, including S5\,1044+71, using log-parabola model was carried out to constrain the spectral parameters of the sources. The study revealed that the emission region in the jet exhibits a reduced radius during the high state, whereas the magnetic field strength increases during the low state \citep{2024ApJ...962...22Z}. Although many studies have been conducted, the underlying cause of blazar variability as well as the location and origin of the $\gamma$-ray flares is still not completely understood.
In this work, we present a long-term multi-wavelength study of the blazar S5\,1044+71, utilizing observations from the \emph{Fermi}-LAT, Swift-XRT and Swift-UVOT. Specifically, we focus on the individual outburst phases recorded between 2008 August and 2024 March. While some outburst phases of S5\,1044+71 have been previously studied, a comprehensive analysis covering all its outburst phases, short-timescale characteristics, and spectral evolution has not been presented in the literature. Our study also aims to investigate potential correlations between flux variations among different wavebands.
Additionally, a detailed broadband SED modeling is carried out in different flux states of the source in order to constrain the underlying emission processes. The framework of this paper is as follows. In section\,\ref{sec:2}, we provide a brief description of the observations and data reduction methods. The results of the muti-wavelength variability of S5\,716+714 are presented in section\,\ref{sec:3}. The findings of the correlation study across different wavebands and the broadband spectral modelling are discussed in subsections\,\ref{sec:3.6} and \ref{sec:3.7}, respectively. Finally, section\,\ref{sec:4} provides the summary and discussion. Throughout this work, we used the flat $\Lambda$ cold dark matter ($\Lambda$CDM) cosmology with $H_0 = 69.6\,\mathrm{km\,s^{-1}\,Mpc^{-1}}$, $\Omega_M = 0.29$, $\Omega_k = 0$, and $\Omega_\Lambda = 0.71$. 

\section{OBSERVATIONS AND DATA ANALYSIS}
\label{sec:2}
In order to perform the multi-wavelength analysis of the source S5\,1044+71, we utilized publicly available data across the $\gamma$-ray, X-ray, ultraviolet (UV), and optical bands, covering a period of approximately 16 years, from 2008 August to 2024 February.
\subsection{\emph{Fermi}-LAT Observations}
\label{sec:2.1}
The \emph{Fermi}-LAT (\emph{Fermi}-Large Area Telescope) is a pair-conversion $\gamma$-ray telescope sensitive to photon energies ranging from 20\,MeV to $>$ 1\,TeV. It surveys the entire sky approximately every 3 hours, with a field of view (FoV) of $\sim$ 2.4 sr \citep{2009ApJ...697.1071A}. This setup is well-suited for observing the short-term evolution of $\gamma$-ray point sources. Utilizing the \emph{Fermi}-LAT public data server,\footnote{\url{https://fermi.gsfc.nasa.gov/cgi-bin/ssc/LAT/LATDataQuery.cgi}} we collected the $\gamma$-ray data for S5\,1044+71 from 2008 August 04 to 2024 February 05 (54682–60350 MJD), covering the energy range of 0.1–300 GeV. 

We analyze the data using the LAT analysis software \texttt{FERMITOOLS} version 2.2.0, following the standard procedures outlined in the \emph{Fermi}-LAT documentation\footnote{\url{https://fermi.gsfc.nasa.gov/ssc/data/analysis/documentation/}}. The analysis is carried out using the \texttt{P8R3\_SOURCE\_V3} instrument response function (IRF). Photons are selected within a circular region of interest (ROI) with a 15$^\circ$ radius, centered at the location of S5\,1044+71. Only photon-like events classified as \texttt{evclass} = 128 and \texttt{evtype} = 3 are selected. A zenith angle cut of $< 90^\circ$ is applied to the data to reduce contamination from background $\gamma$-rays originating from the bright Earth's limb. The good time intervals (GTIs), when the satellite was in standard data-taking mode, were selected using the filter expression \texttt{(DATA\_QUAL $>$ 0) \&\& (LAT\_CONFIG == 1)}. We modeled the Galactic diffuse emission and the isotropic emission components with \texttt{gll\_iem\_v07.fits} and \texttt{iso\_P8R3\_SOURCE\_V3\_v1.txt}, respectively. Utilizing the  \emph{Fermi}-LAT 4FGL catalog \citep{2020ApJS..247...33A}, we generated an XML model file encompassing all sources within the ROI.
During the analysis, the source parameters were left free
within the ROI and constrained to their 4FGL catalog
values outside ROI. To assess the significance of the detection, we used the maximum likelihood (ML) test statistic, defined as TS $= 2\, \Delta \log(\mathcal{L})$, where $\mathcal{L}$ represents the likelihood function comparing models with and without a $\gamma$-ray point source at the source location. During the likelihood analysis, we froze the spectral parameters for the background sources with test statistic TS $<$ 25 ($\sim$5$\sigma$ detection), and the output model file is then used for the generation of lightcurves and spectra for the source. We considered source detection only when TS $>$ 4, which corresponds to $\sim$2$\sigma$ detection \citep{1996ApJ...461..396M}. 

\subsection{Swift-XRT Observations}
\label{sec:2.2}
In this work, the X-ray data were collected by the X-Ray Telescope (Swift-XRT), on board  the \textit{Neil Gehrels Swift Observatory} \citep{2004ApJ...611.1005G}, in the energy range of 0.3–10.0 KeV. During the time period between 54682 to 60350 MJD, a total of 127 Swift observations were available for the source S5\,1044+71. The Swift-XRT lightcurve is generated such that each observation ID corresponds to a single data point on the X-ray lightcurve. Moreover, for each $\gamma$-ray flare of the source, nearly simultaneous observations are also available in the X-ray lightcurve.

For the analysis, we followed the standard procedure provided by the instrumentation pipeline. We used only photon-counting (PC) mode data for the lightcurve and spectral analysis. The data were processed using the \texttt{xrtpipeline} version 0.13.7, incorporating the latest calibration database (CALDB: version 20190910) and response files from \texttt{HEASOFT} version 6.31.1. To generate the energy spectra, we added the calibrated and cleaned event files. The source spectra were extracted from a circular region with a radius of 20 pixels (1 pixel $\sim $2.36 arsc), centered on the source, whereas the background spectra were extracted from a region with a radius of 50 pixel, located away from the source, respectively. The exposure maps were merged using the \texttt{ximage} task, while ancillary response files (ARFs) were generated using the task \texttt{xrtmkarf}. The corresponding response matrix files (RMFs) were retrieved from CALDB. Both the RMFs and ARFs were combined together with the source and background files using the tool \texttt{grppha}, such that each bin contains a minimum of 20 counts. Since the count rate for all the observations were significantly below 0.5 counts s$^{-1}$, pileup corrections were deemed unnecessary. The data were fitted with basic power-law (PL) and log-parabola (LP) models, using the \texttt{XSPEC}, version - 12.13.0 \citep{1996ASPC..101...17A}. During the fitting, the neutral hydrogen column density ($n_H$) were set to $n_H = 3.92 \times 10^{20} \, \text{cm}^{-2}$ \citep{2005yCat.8076....0K}.

\subsection{Swift-UVOT Observations}
\label{sec:2.3}
In addition to the X-ray data, Swift also provides optical/UV data through Swift-UVOT \citep{2005SSRv..120...95R}, which observes in optical and UV bands. The data obtained for S5\,1044+71 were analysed following the standard procedure using \texttt{HEASOFT} package (version-6.31.1). The \texttt{UVOTSOURCE} task within the \texttt{HEASOFT} package were employed to process the images. The observations include all the six filters: v, b, u and w1, m2, w2. Multiple images from the various filters were combined using the \texttt{UVOTIMSUM} tool. We extracted the source counts from a circular region of 5 arcsecond radius centered on the source location, while the background counts were taken from a nearby source-free circular region with a radius of 10 arcsecond. The observed fluxes were corrected for Galactic extinction using $\rm E(B - V) = 0.0535$ and $\rm R_V = \rm A_V / E(B - V) = 3.1$, following \cite{2011ApJ...737..103S}.

\section{RESULTS}
\label{sec:3}
\subsection{Long-term $\gamma-$ray temporal analysis}
\label{sec:3.1}
\begin{figure*}
    \centering
    \includegraphics[width=\textwidth,height=0.4\textheight]{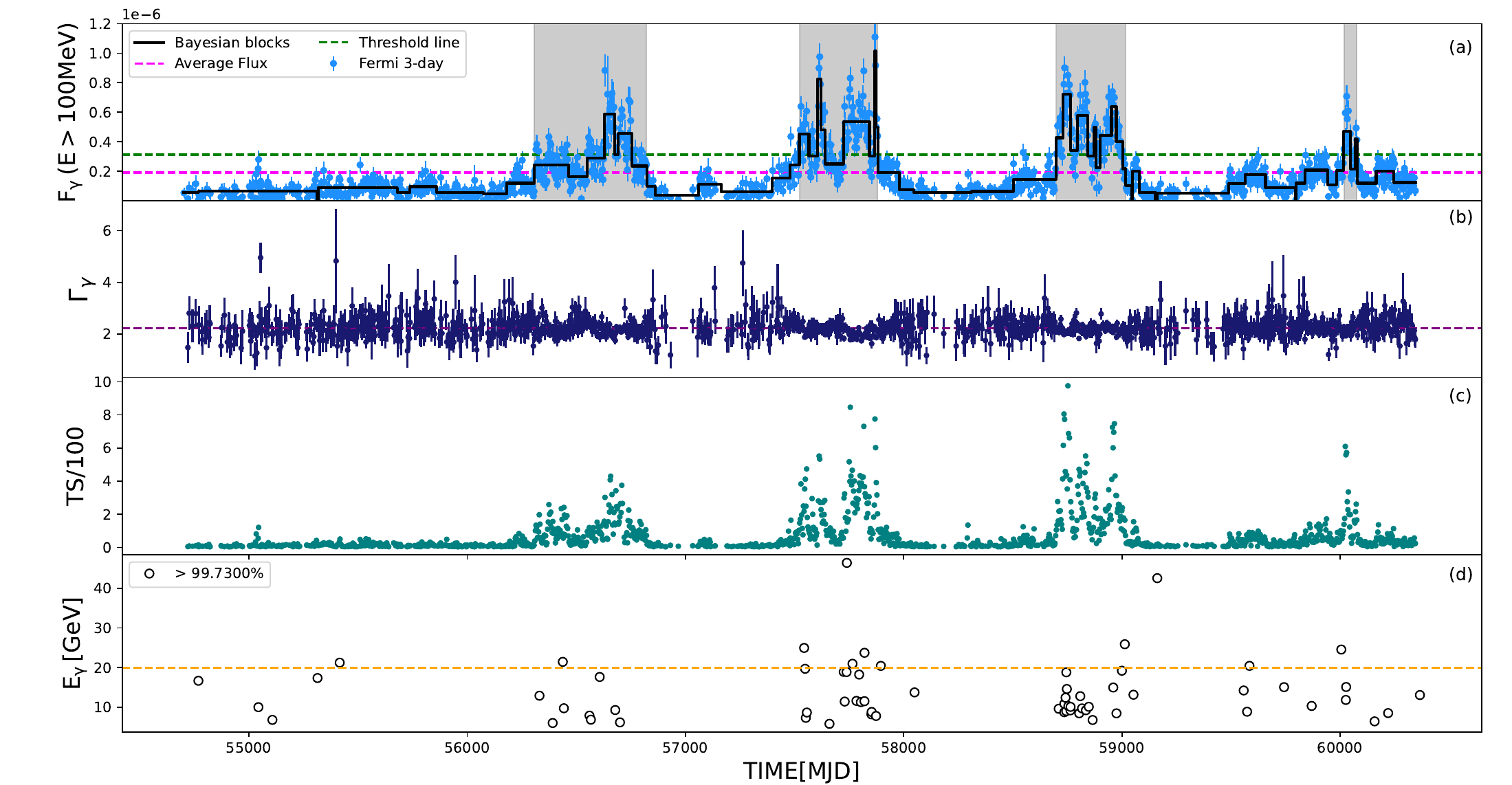}
  \caption{(a) Fermi-LAT lightcurve of S5\,1044+71 from August 2008 to February 2024 at flux $\text{F}_{\gamma}$ (E $>$100\,MeV), with 3-day binning, in units of $\text{ph}\,\text{cm}^{-2}\,\text{s}^{-1}$. The grey vertical strips indicate the time intervals selected for the flux distribution study. (b) $\gamma$-ray spectral-index ($\Gamma_{\gamma}$) as a function of time, with the horizontal dashed purple line representing the average value of index. (c) TS values ($>$ 4) for each time bin. (d) Arrival times and energies of $\gamma$-ray photons E$_{\gamma}$, in units of GeV, with significance level above $3\sigma$. 
  All photons above yellow dashed line have E$_{\gamma}>20$\,GeV.}
    \label{fig: ph_ts}
\end{figure*}

The 3-day binned $\gamma-$ray lightcurve of S5\,104+71 above 100 MeV, from 2008 August to 2024 February is shown in Fig.\,\ref{fig: ph_ts}(a). All flux points displayed in the 3-day binned $\gamma-$ray lightcurve are obtained with TS $>$ 4 (above 2$\sigma$ detection). During the first four years of LAT observations of S5\,1044+71, the source  remained largely in a steady state, with a mean flux of $ \rm (0.78\pm 0.37)\times10^{-7}\,ph\,cm^{-2}\,s^{-1}$. Following this period, the source exhibited four major $\gamma-$ray outburst phases (or flare epochs), showing significant flux variability with flux reaching upto $ \rm (1.11\pm 0.09)\times10^{-6}\,ph\,cm^{-2}\,s^{-1}$ on 2017–04–25 (57868.5 MJD). In the 3-day binned $\gamma$-ray lightcurve, the flare epochs were identified by using the Bayesian block algorithm \citep{2013ApJ...764..167S}. Here, the Bayesian block algorithm was directly adopted from \texttt{astropy.stats} with a false alarm rate parameter of $p_{0} <$ 0.001 (corresponding to $>$3$\sigma$). In order to determine the time range of the flare epochs, we apply Ivan Kramarenko's algorithm and consider monotonically decreasing sets of adjacent blocks. Following \cite{2020ApJ...904...67G,2023MNRAS.521.3451D}, the algorithm uses an iterative approach to divide the lightcurve data points into two sets: a low-state set and a high-state or ``anti-set". The threshold value  for the flaring state is obtained as $ \rm 3.12\times10^{-7}\,ph\,cm^{-2}\,s^{-1}$ (dashed green line in Fig.\,\ref{fig: ph_ts}(a)). It is calculated by using the expression $\langle F \rangle + 2 \times F_{\text{Disp}}$, where $\langle F \rangle$ is the mean flux of the lightcurve, and $F_{\text{Disp}}$ represents the true dispersion of the low state group. 
Therefore, the determined flare epoch durations are represented as: S1 (56306.5 – 56821.5 MJD), S2 (57523.5 – 57880.5 MJD), S3 (58699.5 – 59016.5 MJD) and S4 (60017.5 – 60076.5 MJD), and shown by shaded strips in Fig.\,\ref{fig: ph_ts}(a). In each epoch, the average flux exceeds the mean flux value of the entire
 $\gamma$-ray lightcurve, $ \rm (1.91 \pm 0.4) \times 10^{-7}\,ph\,cm^{-2}\,s^{-1}$, (Fig.\,\ref{fig: ph_ts}(a), dashed magenta line).

\begin{figure*} 
    \centering
    \includegraphics[width=\textwidth,height=0.3\textheight]{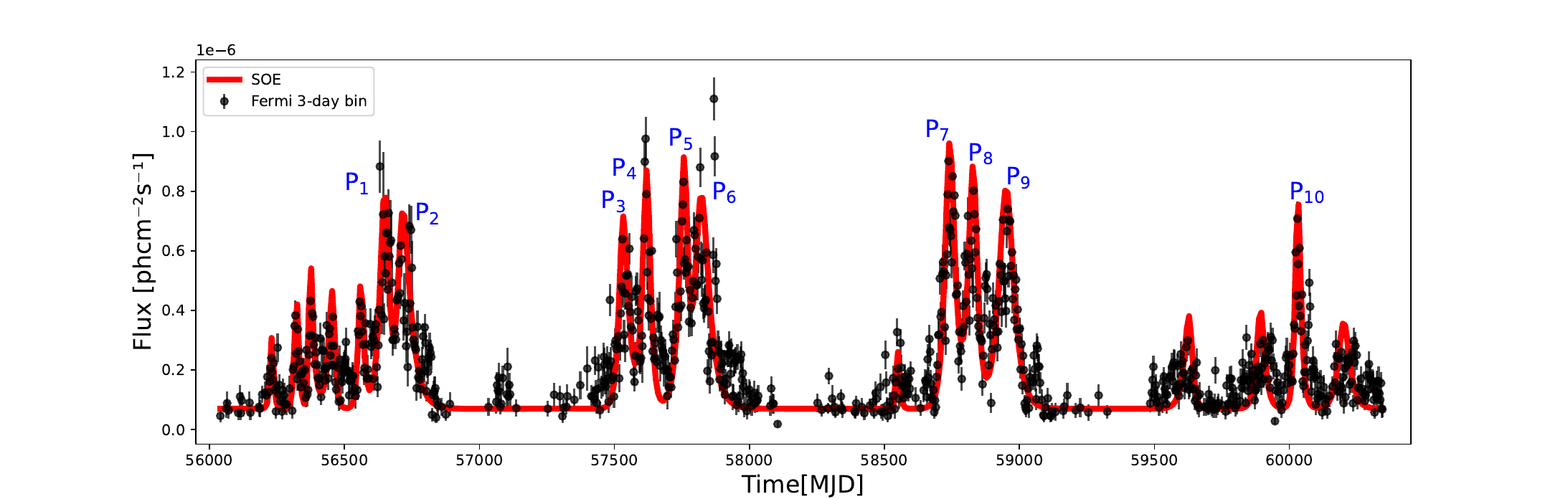}
    \caption{3-day binned $\gamma$-ray lightcurves of S5\,1044+71 fitted with the SOE function defined in equation\,(\ref{eq:SOE}).}
    \label{fig: flux_plot}
\end{figure*}

The $\gamma$-ray lightcurve of S5\,1044+71 exhibits large flux variations, with each flare epoch consisting of multiple peaks. We study the temporal characteristics of the flare epochs and probe the evolution of the peak components during 55560.5 – 60344.5 MJD. We adopt the following sum of exponential (SOE) function to fit the temporal profile of the $\gamma$-ray lightcurve to compute the rise ($T_{r}$) and decay ($T_{d}$) times of each peak component \citep{2010ApJ...722..520A}
\begin{equation}
F(t) = F_{b} + 2 \sum_{i=1}^{n} F_{0,i} \left[ \exp\left( \frac{t_{0,i} - t}{T_{r,i}} \right) + \exp\left( \frac{t - t_{0,i}}{T_{d,i}} \right) \right]^{-1},
  \label{eq:SOE}
\end{equation}
where $i$ runs over the $n$ major peaks, $F_{0}$ represents the photon flux at time $t_{0}$, and $F_{b}$ denotes the baseline flux. For better photon statistics, we consider only flux points with  ratio of flux and error in flux $\rm F/\delta F>2$, and test statistics value TS $>$ 9 (significance $>$ 3$\sigma$). The fitted SOE profile for the 3-day binned $\gamma$-ray lightcurves is shown in Fig.\,\ref{fig: flux_plot}. The time $T_{m}$ of maximum of a flare is given by \citep{2015ApJ...807...79H}
\begin{equation}
T_{m} = T_0 + \frac{T_r T_d}{T_r + T_d} \ln \left( \frac{T_d}{T_r} \right),
\end{equation}
where $T_{m}$ is equal to $T_0$ when $T_r$ = $T_d$. The symmetry of the flare components were obtained by
\begin{equation}
\xi = \frac{T_r -  T_d}{T_r + T_d},
\end{equation}
where \(|\xi| < 1\). 
For the components of 3-day binned $\gamma$-ray lightcurve with peak flux exceeding $3.12\times10^{-7}\,\mathrm{ph\,cm^{-2}\,s^{-1}}$ (threshold line) and  $\xi / \xi_{err}\geq 2 $, the best fit parameters and their corresponding uncertainties  are presented in Table\,\ref{tab: rt_dt}. 
  Further, flares with $\xi<0$ are known as fast-rise exponential-decay (FRED) type flares and are commonly observed in $\gamma$-ray lightcurves. These flares suggest that the injection of energetic particles occurs on timescales shorter than their subsequent cooling through radiative processes like inverse Compton (IC) scattering or synchrotron emission \citep{2019ApJ...877...39M,2024JHEAp..42..115T}. However, in our analysis, we identified several flares with $\xi < 0$, with most flares displaying significant symmetry (Table\,\ref{tab: rt_dt}).
\begin{table*}
    \centering
\caption{Rise and Decay times for the 3-day bin $\gamma$-ray lightcurve integrated over the energy range 0.1–300 GeV during (55560.5 – 60344.5 MJD). Column 1: Individual peaks; 2, 3: Peak time (MJD), Peak flux ($10^{-7} \text{ph}\,\text{cm}^{-2}\,\text{s}^{-1}$), 4, 5: Rise time, decay time, of the components; 6: Asymmetry parameter and 7: $\xi/\xi_{err}$.}
\renewcommand{\arraystretch}{1.2} 
\begin{tabular}{cccccccc} \\
\hline
\hline
    Peak & $\mathrm{t_0}$ & $\mathrm{F_0}$ &$\mathrm{T_r}$ & $\mathrm{T_d}$ &$\xi$ & $\xi/\xi_{err}$\\
               & (MJD) & ($10^{-7} \text{ph}\,\text{cm}^{-2}\,\text{s}^{-1}$) &(days) &(days)\\ 
       \hline 
       (1)&(2)&(3)&(4)&(5)&(6)&(7)\\
      $\mathrm{P_1}$ &56650&7.24& $15.37\pm0.59$ & $12.85\pm 1.10$ & $0.08 \pm 0.04$ &2.00\\
      $\mathrm{P_2}$ &56710&6.09& $12.48\pm 1.05$ & $29.16\pm 0.86$ &$-0.40\pm 0.03$ &13.34\\ 
      $\mathrm{P_3}$ &57530&6.32& $11.71\pm 0.56$ & $17.69\pm 0.77$ &$ -0.20\pm 0.03$ &6.67\\
      $\mathrm{P_4}$ &57620& 7.91& $13.30\pm 0.61$ & $11.84\pm 0.44$ &$ 0.05\pm 0.02$ &2.50\\
      $\mathrm{P_5}$ &57762& 7.57& $17.93\pm 0.63$ & $8.96\pm 0.74$ & $0.33\pm 0.04$ &8.25\\
      $\mathrm{P_6}$&57817& 7.01& $17.49\pm1.11$ & $27.64\pm 0.72$ &$-0.22 \pm 0.03$&7.33 \\
      $\mathrm{P_7}$ &58741& 8.89& $14.59\pm0.44$ & $16.73\pm0.60$ &$-0.06 \pm 0.02$&3.00 \\ 
      $\mathrm{P_8}$ &58830& 7.99& $16.03\pm0.63$ &$13.00\pm0.48$ & $0.10 \pm 0.03$ &3.33\\
      $\mathrm{P_9}$ &58950& 7.40& $22.95\pm0.71$ & $25.02\pm0.59$ &$-0.04 \pm 0.02$ &2.00\\ 
      $\mathrm{P_{10}}$ &60030& 7.05&$7.79\pm0.38$& $10.55\pm0.40$  &$ -0.15\pm 0.03$& 5.00\\
    \hline
    \hline   
\end{tabular}
\label{tab: rt_dt}
\end{table*}

To characterize the variability during different flaring epochs, we estimated the  variability timescale of the source. We scanned the $\gamma$-ray lightcurve to determine the shortest flux doubling timescale by using the equation 
\begin{equation}
F(t) = F(t_0) \, 2^{\frac{t - t_0}{\tau}},
\end{equation}
where $F(t)$ and $F(t_0)$ are the flux values measured at times $t$ and $t_0$, with $\tau$ representing the characteristic flux doubling or halving time. On applying the condition that the significance of the difference in flux at time $t$ and $t_0$ is $\geq 3\sigma$ \citep{2011A&A...530A..77F}, we found the shortest variability time scale of $4.5$\,hr. The flux variability timescale can be used to constrain the size and location of the $\gamma$-ray emission region near the central black hole.

\subsection{Search for highest energy photon}
\label{sec:3.2}
We investigate the emission of high-energy $\gamma$-ray photons having energy $E_\gamma>$ 20\,GeV. The observation of such photons is generally difficult to explain if the emission is assumed to originate only within the inner BLR, as photon–photon pair production would likely inhibit the escape of high-energy photons \citep{2003PThPS.151..186D,2008ApJ...688..148L,2009ApJ...703.1168B}. In order to search for high-energy photons associated with S5\,1044+71, we utilized the \texttt{gtsrcprob} tool provided by the \emph{Fermi} science package. This tool calculates the probability for the detected photons to be associated with the source ROI. Fig.\,\ref{fig: ph_ts}(d) shows the lightcurve of the high-energy photons, emitted over a period of around 15 years, having $\geq$ 99.73 per cent  probability of originating from the source. The two highest-energy photons observed, with measured energies of 46.4\,GeV at 57739.6 MJD and 42.5\,GeV at 59161.9 MJD, each with a significance level of $>3\sigma$, are displayed in Fig.\,\ref{fig: ph_ts}(d). 
We note that both high-energy photons were detected outside the peak flaring activity of the source. Although the 46.4 GeV photon falls within the broader temporal interval associated with the second flaring epoch S2 (57523.5 – 57880.5 MJD), it is clearly offset from the flare maximum and coincides with a relatively low $\gamma$-ray flux state within that epoch. In general, enhanced $\gamma$-ray activity increases the probability of detecting high-energy photons; however, their detection during low-flux states indicates that the conditions required for the production and escape of the highest-energy photons are not necessarily tied to peak flaring activity. The detection of $\sim$40–50 GeV photons from S5 1044+71 can provide useful constraints on the physical conditions and location of the $\gamma$-ray emission region. The presence of such photons during low-activity phases suggests that they may originate from emission zones characterized by reduced soft-photon densities or enhanced Doppler boosting, which lowers the internal $\gamma$--$\gamma$ opacity and facilitates photon escape \citep[e.g.,][]{2010ApJ...717L.118P}. In addition, attenuation by the extragalactic background light (EBL) at these energies is expected to be moderate \citep{2010arXiv1005.2626F}, whereas absorption by local photon fields, particularly those associated with the BLR, can be significant if the emission region is located within the BLR. The detection of such high-energy photons therefore favors an origin at or beyond the BLR, or physical conditions that allow $\gamma$-rays to escape even during low-flux states \citep[e.g.,][]{2013MNRAS.435L..24T}. 


\subsection{$\gamma$-ray flux-index correlation}
\label{sec:3.3}
The flux versus index correlation is studied to find the association of index hardening or softening with different flux states. We used the Spearman rank correlation method to compute the correlation coefficient  ($r_s$) and the null hypothesis probability ($p$) between the flux and index values for the selected epochs. The correlation analysis was performed with flux points having test statistic $\rm TS>9$ (significance $>3\sigma$) and signal-to-noise ratio $\rm F/\delta F>2$. The results of correlation analysis are provided in Table\,\ref{tab: spearman_parameters}. We observed that epochs S1, S3 and S4 do not exhibit spectral hardening, while epoch S2 is characterized by a weak spectral hardening with $r_s = -0.28$ and $p=2.36\times10^{-3}$. Moreover, a slight indication of mild negative correlation was observed during epoch S4, with a generally high scatter across all cases. A similar type of behavior has been noticed from several blazars, for example PKS 1510-089 \citep{2010ApJ...716...30A} and 3C 454.3 \citep{2010ApJ...721.1383A}.
The presence of a weak “harder-when-brighter” trend during the flare epoch S2 may indicate a temporary enhancement in particle acceleration efficiency or a hardening of the underlying electron energy distribution during this epoch. However, the correlation remains modest and is accompanied by substantial scatter, suggesting that statistical limitations and the possible overlap of multiple emission states likely affect the observed behavior. For the other epochs, the lack of a clear flux–index correlation may result from poorer photon statistics or the coexistence of multiple emission components. Therefore, while the observed behavior is qualitatively consistent with that reported for other blazars, the present data do not permit firm conclusions about the presence of distinct flaring or acceleration mechanisms across the different epochs.
\begin{table}[ht]
\centering
\caption{Spearman rank correlation coefficient ($r_s$) and p-value for the flaring epochs S1 to S4.}
\label{tab: spearman_parameters}
\renewcommand{\arraystretch}{1.4} 
\begin{tabular}{lrc}
\hline
\hline
Flare epoch & $\rm r_s$& p-value \\
\hline
S1 & $- 0.03$ &$6.63\times 10^{-1}$  \\
S2 & $- 0.28$ & $2.36\times 10^{-3}$ \\
S3 & $ 0.01$ & $8.93\times 10^{-1}$ \\
S4 &  $-0.13$& $5.94\times 10^{-1}$ \\
\hline
\hline
\end{tabular}
\label{tab: spearman}
\end{table}

\subsection{Flux distribution}
\label{sec:3.4}
The log-normal variability observed in blazar lightcurves is indeed an intriguing phenomenon. Many blazars exhibit typical log-normal flux distribution across different energy bands and timescales \citep{2010A&A...520A..83H,2018RAA....18..141S}. Analysis of \emph{Fermi}-LAT lightcurves also indicates a predominant adherence of blazars to the log-normal flux distribution \citep{2015ApJ...810...14A,2018RAA....18..141S,2021MNRAS.502.5245P}. Such distributions can arise from multiplicative processes, such as the integration of numerous mini-jets, where the logarithm of the composite flux follows a normal distribution \citep{2012A&A...548A.123B}. Moreover, introducing a Gaussian perturbation to the particle acceleration timescale can also lead to the log-normal flux distribution in blazar lightcurves \citep{2018MNRAS.480L.116S}. In addition to single log-normal characteristics, flux histograms from certain blazars display a double log-normal distribution pattern \citep{2020Galax...8...66K}. Furthermore, the presence of double log-normal flux distribution in the X-ray emissions of blazars such as Mrk 501 and Mrk 421 has been attributed to the underlying dual Gaussian structure in the spectral index \citep{2020MNRAS.491.1934K}. 

We analyzed the 3-day binned $\gamma$-ray lightcurve of S5\,1044+71, during the flaring epochs S1, S2, S3 and S4, to study the flux distributions. To ensure good photon statistics, we include only flux points with $ \rm F/\delta F >2$. To examine the flux characteristics of the source, we performed the Shapiro–Wilk (SW) test \citep[see, e.g.,][]{shapiro1965analysis,razali2011power}, skewness, and histogram fitting test for all cases. SW test is a statistical tool that can be used to perform a normality test by evaluating the null hypothesis $(H_0)$ that the data are drawn from a normal distribution. The test calculates the null hypothesis probability value (p-value), and if the p-value is $\leq 0.05$, the null hypothesis is rejected, suggesting that the sample deviates from the normal distribution. 
The SW tests for flare epochs S1, S2 and S4 show that the p-values for 3-day binned fluxes in log-scale are greater than 0.05, indicating that the binned flux distributions are log-normal in nature (see Table\,\ref{tab: sw_test}). However, for S3 and the entire 3-day binned lightcurve, the p-values in both linear and log-scale are less than 0.05, suggesting that the binned flux distributions are neither normal nor log-normal. This result is further supported by the log-scale skewness test, where the test statistic values of –3.91 and 2.92, with p-values less than 0.05, indicate statistically significant skewness (see Table\,\ref{tab: sw_test}, last column). 

\begin{figure*} 
    \centering
    \includegraphics[width=0.3\textwidth]{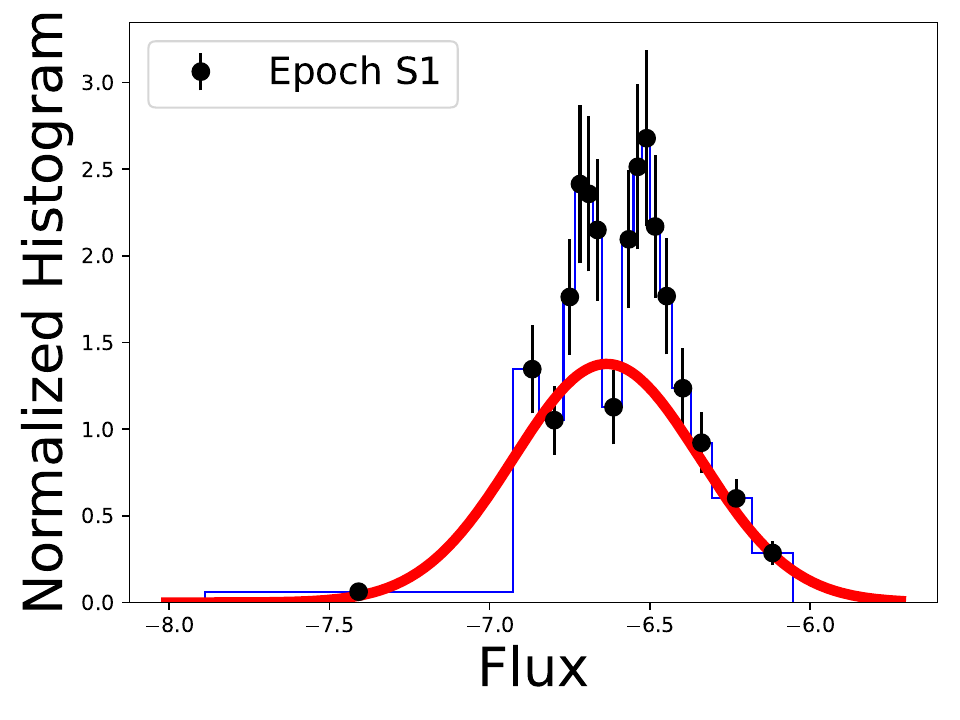}
    \includegraphics[width=0.3\textwidth]{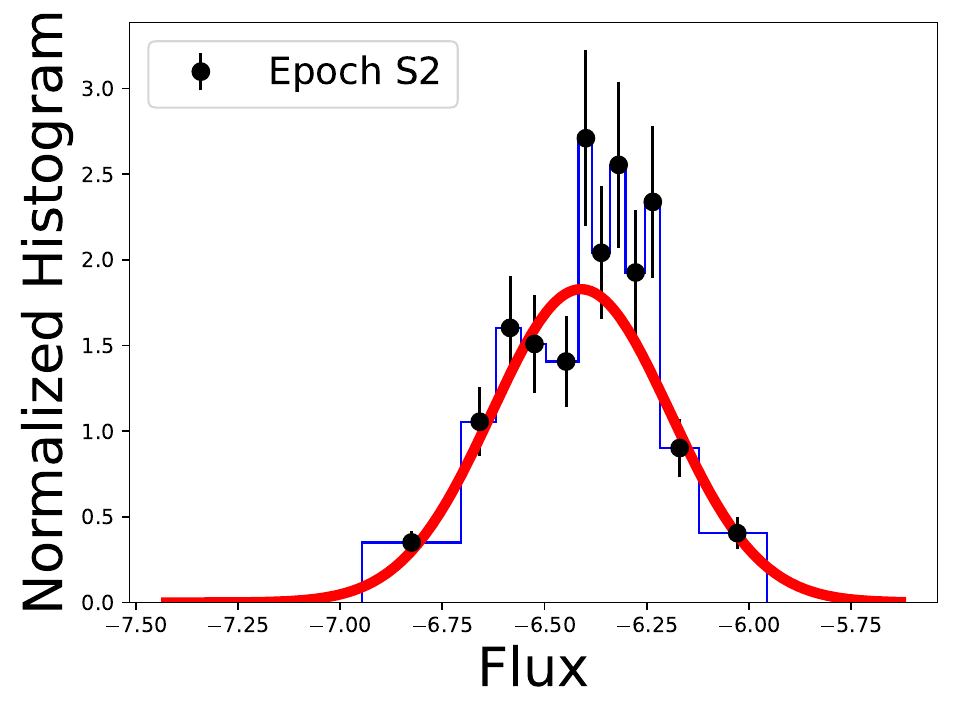}
    \includegraphics[width=0.3\textwidth]{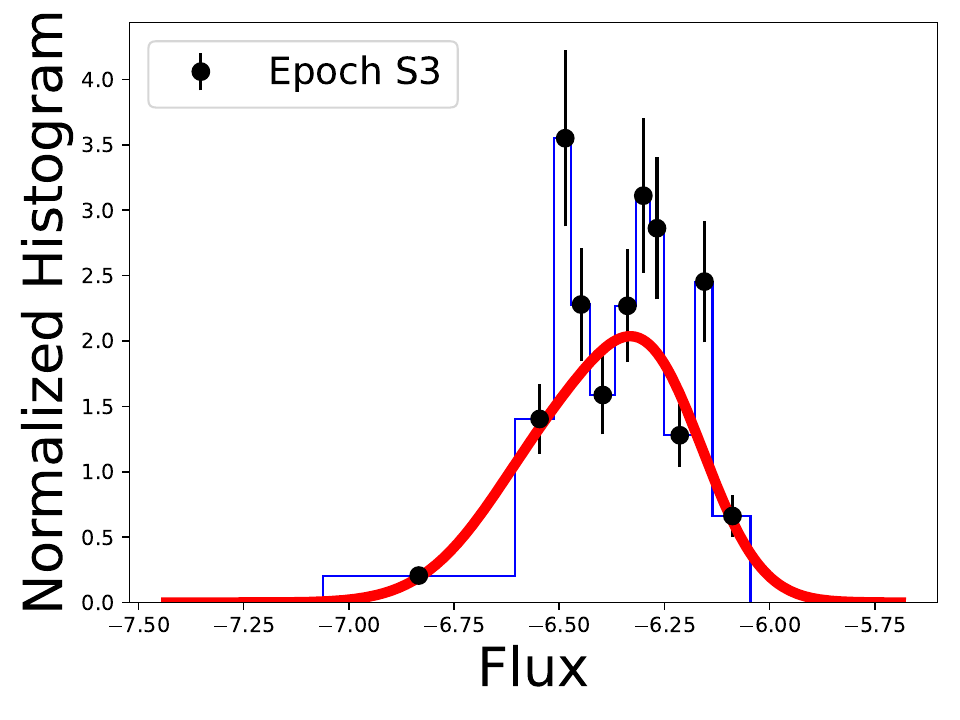}
    \includegraphics[width=0.305\textwidth]{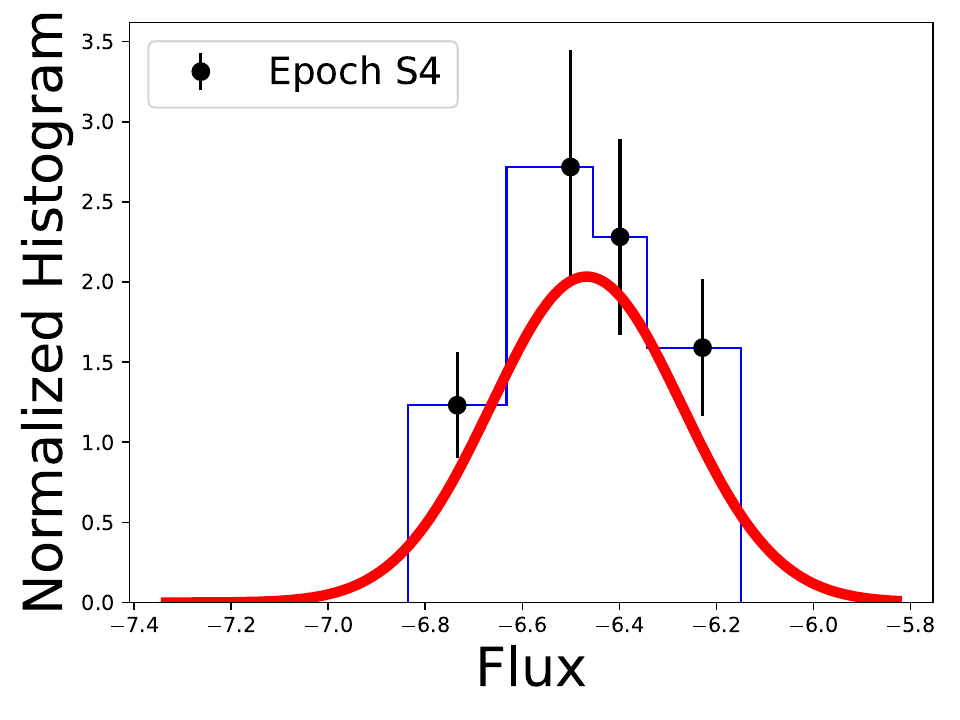}
    \includegraphics[width=0.32\textwidth]{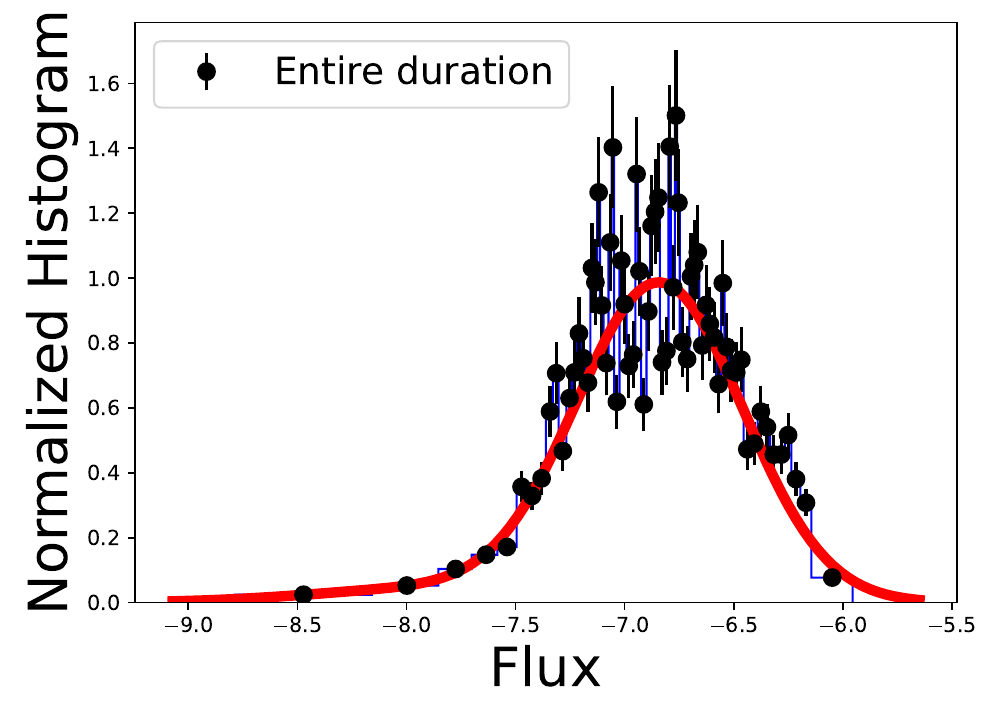}
    	\caption{Flux distribution of S5\,1044+71 in the $\gamma$-ray band. Epochs S1, S2, and S4 are fitted with a Gaussian PDF, whereas epoch S3 and the entire 3-day binned $\gamma$-ray lightcurve are fitted with a double Gaussian PDF, respectively. The flux are in units of $\text{ph}\,\text{cm}^{-2}\,\text{s}^{-1}$.}

    \label{fig:lognr}
\end{figure*}

We constructed histograms of normalized flux counts for the selected epochs and the entire 3-day binned lightcurve in log-scale. The histograms were generated such that each bin contains an equal number of flux points, while allowing the bin width to vary. The normalized flux histograms are plotted in Fig.\,\ref{fig:lognr}. We further analyzed the flux
distributions by fitting their normalized histograms with a probability density function (PDF) 
\begin{equation}
\label{eq:pdf}
    f(x) = \frac{1}{\sqrt{2\pi\sigma^{2}}} e^{-\frac{(x - \mu)^{2}}{2\sigma^{2}}},
\end{equation}
with $\mu$ and $\sigma$ being the centroid and width of the distribution, and a double-PDF function \citep{2016ApJ...822L..13K}
\begin{equation}
\label{eq:dpdf}
    g(x) = \frac{a}{\sqrt{2\pi\sigma_{1}^{2}}} e^{-\frac{(x-\mu_{1})^{2}}{2\sigma_{1}^{2}}} + \frac{(1-a)}{\sqrt{2\pi\sigma_{2}^{2}}} e^{-\frac{(x-\mu_{2})^{2}}{2\sigma_{2}^{2}}}
\end{equation}
where, a is is a mixing fraction that determines the relative contribution of the two components and $\mu_{1}$, $\mu_{2}$ represent the centroids of the distributions with widths $\sigma_{1}$, $\sigma_{2}$, respectively. 
The corresponding log-normal or double log-normal fits to the histograms are shown in Fig.\,\ref{fig:lognr}, and the best-ﬁt parameter values are provided
in Table\,\ref{tab: lognr}. The reduced $\chi^2$ values suggest that the flux histograms from S1, S2 and S4 are log-normal, while the flux histograms from S3 and the entire 3-day binned $\gamma$-ray lightcurve are double log-normal in nature. Due to the small number of data points in S4, this result must be interpreted as tentative. Furthermore, these results are also consistent with the results obtained from SW test statistics. 
\begin{table*} 
	\centering
	\caption{Summary of SW test results for normal and log-normal distributions of flux during different epochs.}
\renewcommand{\arraystretch}{1.4} 
\begin{tabular}{ccclllc}
\hline  
\hline
{Epoch} & {Period}  & {Number of } & {Normal flux} & \multicolumn{2}{c}{Log-normal flux } &  \\ 
&&{Data Points}&SW (p-value)&SW (p-value)& Skewness (p-value)&\\
\hline

S1&$56306.5 - 56821.5$
&$166$&$0.89 (1.53\times10^{-9})$&$0.99 (0.34)$&$0.97 (0.33)$&\\
S2&$57523.5 - 57880.5$&$117$&$0.94 (1.31\times10^{-4})$&$0.98(0.38)$&$-0.78(0.43)$&\\
S3&$58699.5 -59016.5$&$106$&$0.98(4.40\times10^{-3})$&$0.94(1.38\times10^{-4})$&$-3.91(8.9\times10^{-5})$&\\
S4&$60017.5 - 60076.5$&$21$&$0.95(0.43)$&$0.94(0.36)$&$-0.61(0.53)$&\\
Entire duration &$54702.5 - 60344.5$&$1320$&$0.85(4.52\times10^{-28})$&$0.98(1.54\times10^{-8})$&$2.92(3.41\times10^{-3})$&\\
\hline
\hline
\end{tabular}%
\label{tab: sw_test}
\end{table*}
\begin{table*} 
    \centering
    \caption{Best-ﬁtting parameter values obtained by fitting equations (\ref{eq:pdf}) and (\ref{eq:dpdf}) to the logarithm of the fluxes in different selected epochs, respectively.}
    \renewcommand{\arraystretch}{1.3} 
    \begin{tabular}{cccccccccc}
        \hline
        \hline
        {Epoch} & \multicolumn{2}{c}{Log-normal PDF} & \multicolumn{5}{c}{Double Log-normal PDF}&$\chi_{\text{red}}^2$ \\
        & $\mu$ & $\sigma$ &$a$&$\mu_1$ & $\sigma_1$&$\mu_2$ & $\sigma_2$& \\
        \hline
        S1 & $-6.63\pm 0.03$ & $0.28\pm 0.02$ &  &&&&&  $1.28$\\
        S2 &$-6.40\pm 0.02$  &$0.21\pm 0.02$  & & & &&&$1.07$ \\
        S3 &  &  & $0.34$  &$-6.46\pm 0.12$&$0.18\pm 0.02$ &$-6.26\pm 0.07$ &$0.12\pm 0.03$ &$1.68$ \\
        S4 & $-6.44\pm0.04$ & $0.17\pm0.05$ &  & &&&&$0.91$ \\
        Entire duration &  &  & $0.06$ & $-7.87\pm 0.09$&$0.54\pm 0.06$ &$-6.83\pm 0.11$&$0.38\pm 0.03$ &$1.12$ \\
        \hline
        \hline
    \end{tabular}%
  \label{tab: lognr}  
\end{table*}

\subsection{Multi-wavelength lightcurve}
\label{sec:3.5}
\begin{figure*} 
    \centering
\includegraphics[width=\textwidth,height=0.4\textheight]{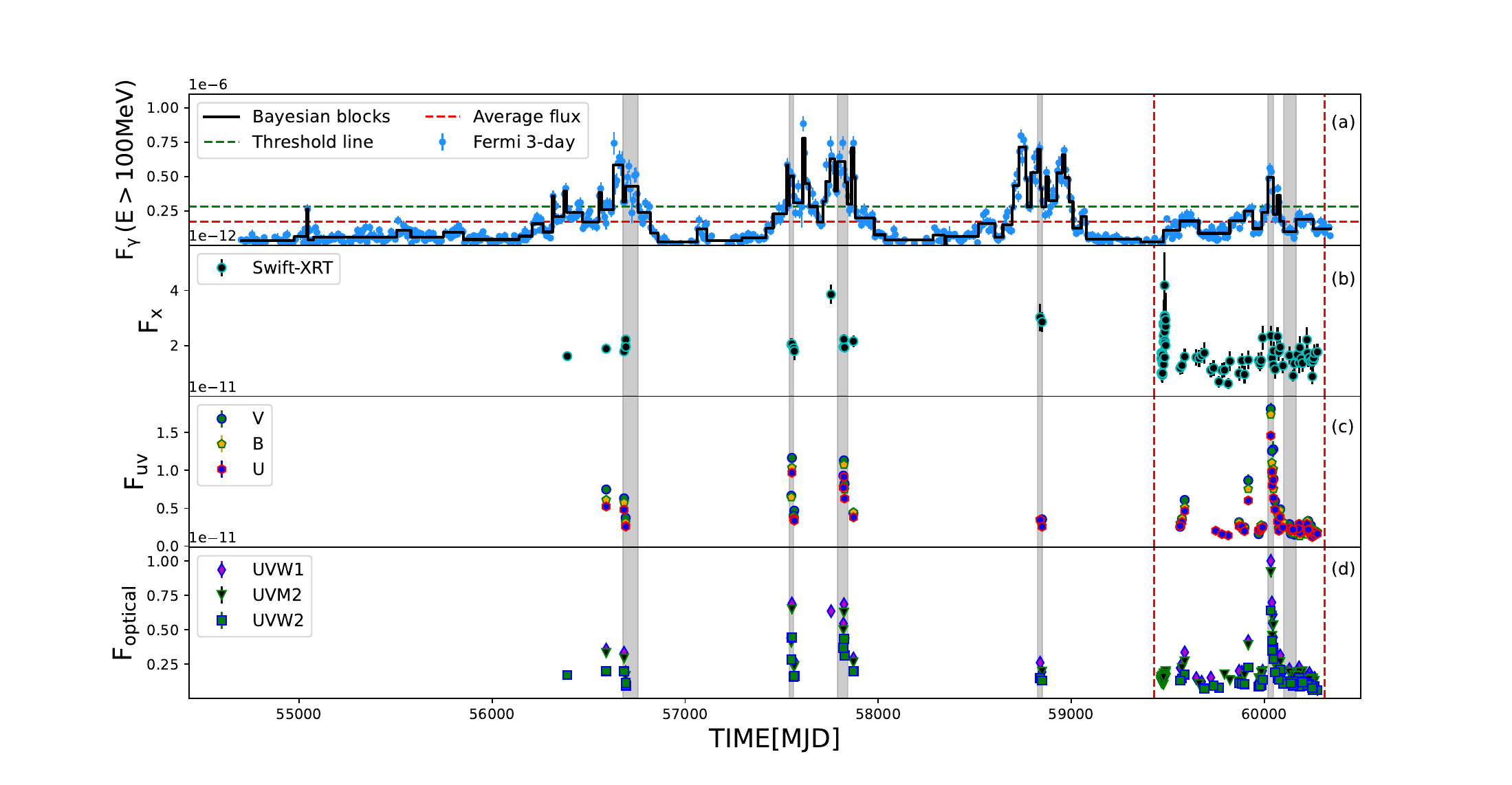}
    \caption{Multi-wavelength lightcurves (MLCs) of S5\,1044+71 obtained by using \textit{Fermi}-LAT and Swift-XRT/UVOT observations. The observations spanned a period from 54702.5 – 60344.5 MJD. Top panel is 7-day binned $\gamma$-ray lightcurve, second, third, and fourth panels are the X-ray, UV, and optical band lightcurves, respectively. The grey vertical stripes indicate the regions where broadband spectral modelling is performed, while the region between the red vertical lines marks the time interval used for cross-correlation analysis among different bands.}
    \label{fig: multlc}
\end{figure*}

\begin{figure} 
    \centering  
    \includegraphics[width=0.45\textwidth]{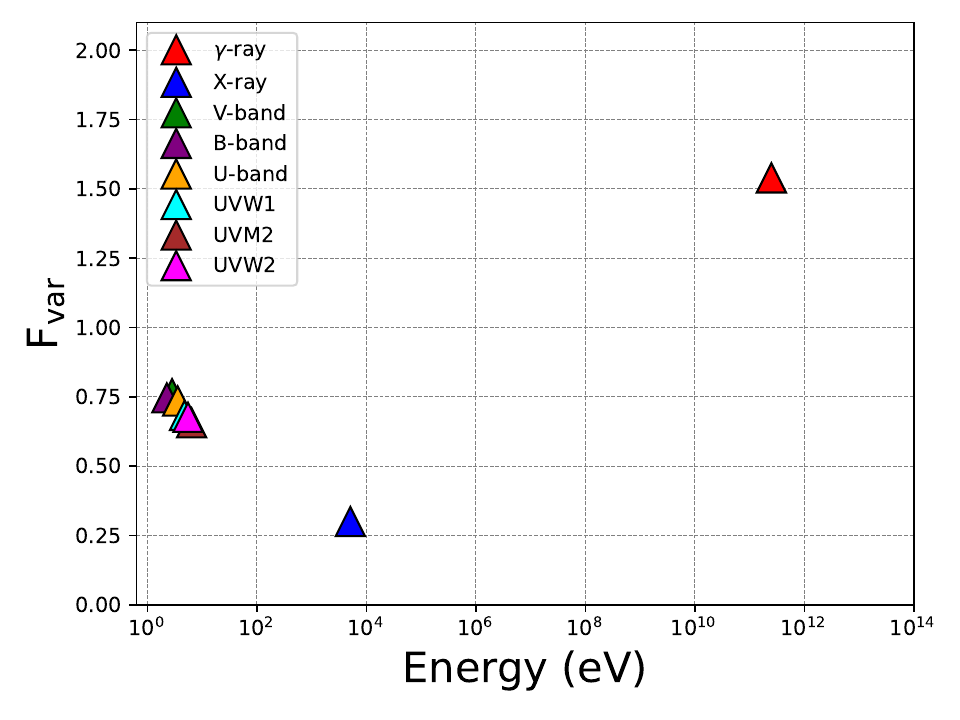}
\caption{Fractional variability amplitude in the $\gamma$-ray, X-ray, optical, and UV bands plotted as a function of energy over the period 54702.5 – 60344.5 MJD.}
    \label{fig: fvar}
\end{figure}

We generated the multi-wavelength lightcurve (MLC) of S5\,1044+71 over a period of approximately 16 years, from  2008 August 08, to 2024, February 04 (54702.5 – 60344.5 MJD). Fig.\,\ref{fig: multlc} shows the multi-wavelength lightcurve of the source constructed by combining the data obtained across $\gamma$-ray, X-ray, and optical/UV bands. Fig.\,\ref{fig: multlc}(a) shows the 7-day binned $\gamma$-ray lightcurve, obtained by integrating over the energy range 0.1 – 300 GeV, displaying multiple quiescent and flaring periods. Swift-XRT/UVOT made a total of 127 observations of the source with X-ray and optical/UV data analyzed by following the analysis procedure described in section\,\ref{sec:2}. Fig.\,\ref{fig: multlc}(b), \ref{fig: multlc}(c), and \ref{fig: multlc}(d) display the X-ray, UV, and optical lightcurves of the source, respectively. The multiplot illustrates the simultaneous flux variations across different energy bands. We observe that the $\gamma$-ray and optical/UV fluxes exhibit notable variations, whereas the X-ray flux shows only minimal fluctuations. 
We computed the variability of the source in each energy band by calculated the fractional RMS variability amplitude ($F_{var}$) using the equation \citep{2003MNRAS.345.1271V}
\begin{equation}
\centering 
    F_{var} = \sqrt{\frac{S^{2} - \langle \sigma_{err}^{2} \rangle}{{\langle f \rangle}^{2}}}
    \label{eq:var_eqn},
\end{equation}
where \text{$\langle f \rangle$} is mean flux, $S^{2}$ is the variance of the total number of flux points in the lightcurve  and  $\langle \sigma_{err}^{2} \rangle$ is their mean square error. The uncertainty in $F_{var}$ is given by \citep{2003MNRAS.345.1271V}
\begin{equation}
    \label{eq:_eqn}
    (F_{\text{var}})_{\text{err}} = \sqrt{\frac{1}{2N} \left(\frac{\langle \sigma_{\text{err}}^{2} \rangle}{F_{\text{var}} \langle f \rangle^{2}}\right)^2 + \frac{1}{N}\frac{\langle \sigma_{\text{err}}^{2} \rangle}{\langle f \rangle^{2}}}
\end{equation}
where, $N$ is the number of data points in the lightcurve. The obtained $F_{var}$ values are presented in Table\,\ref{tab: frac_var},  and the plot of $F_{var}$ versus energy is shown in Fig.\,\ref{fig: fvar}. The $F_{var}$ shows an overall increasing trend from the UV/optical to the $\gamma$-ray band, with a noticeable dip in the X-ray range and a maximum observed in the $\gamma$-ray band. 
This trend in $F_{var}$ can be potentially linked to the energy of relativistic electrons \citep{2016ApJ...819..156B,2016A&A...590A..61C}, and is also evident from the broadband SED of blazars. The higher energy electrons responsible for $\gamma$-ray emission, through the inverse Compton process, cool faster than low-energy electrons resulting in faster variability in the $\gamma$-ray band.
\begin{table}
\centering
  \caption{Fractional variability amplitude obtained in $\gamma$-ray, X-ray, optical, and UV energy bands during 54702.5 – 60344.5 MJD.}
\renewcommand{\arraystretch}{1.3} 
\begin{tabular}{l c}
\hline 
\hline
Energy band  & $\rm F_{var}$ \\
\hline
$\gamma$-ray (\mbox{0.1--300 GeV}) & $1.53\pm 0.01$\\
X-ray (0.3--10 keV) & $0.29 \pm 0.02$\\
UVOT band-W2 & $0.67 \pm 0.09$ \\
UVOT band-M2 & $0.65 \pm 0.08$\\
UVOT band-W1 & $0.68 \pm 0.01$\\
UVOT band-U &$ 0.73 \pm 0.09$\\
UVOT band-B & $0.74 \pm 0.01$\\
UVOT band-V & $0.76 \pm 0.01$\\
\hline
\hline
\end{tabular}
\label{tab: frac_var}
\end{table}

\subsection{Cross-correlation analysis}
\label{sec:3.6}
The investigation of cross-correlation across different energy bands can reveal information about the radiation mechanism and the location of emission region in different bands \citep{2014MNRAS.445..428M,2015MNRAS.452.1280R,2018MNRAS.480.5517L}. It can also be used to determine whether emissions in different bands are co-spatial or arise from distinct regions within the jet \citep{2016MNRAS.456..171R}. We carried a cross-correlation analysis on $\gamma$-ray, X-ray, and UV-optical band lightcurves using the discrete correlation function (DCF). This method is ideal for handling discrete time series data with uneven spacing, particularly for X-ray and UV-optical bands. The DCF calculates correlation coefficients directly without the use of interpolation across unevenly sampled data. The unbinned discrete correlation function (UDCF) for two data sets is defined by \citep{1988ApJ...333..646E}:
\begin{equation}
\label{eq:eldsn}
\text{UDCF}_{ij} = \frac{(a_i - \langle a \rangle)(b_j - \langle b \rangle)}{\sqrt{(\sigma_a^2 - e_a^2)(\sigma_b^2 - e_b^2)}}
\end{equation}
where, $a_i$ and $b_i$ are the two discrete time series, $\sigma$ and $e$ are the standard deviation and measurement error associated with each set. Here, each value of $\text{UDCF}_{ij}$ for the pair ($a_i,b_j$) is associated with the time delay $\Delta t_{ij}=t_j-t_i$. By averaging equation~(\ref{eq:eldsn}) over $M$ number of pairs for which
$\tau-\Delta\tau/2\leq\Delta t_{ij}  (=t_j-t_i)\leq\tau+\Delta\tau/2$, we obtain discrete correlation function given by
\begin{equation}
    \text{DCF}(\tau) = \frac{1}{M} \sum_{ij} \text{UDCF}_{ij} \pm \sigma_{\text{DCF}} (\tau),
\end{equation}
where, $\sigma_{\text{DCF}} (\tau)$ is the associated error at time lag $\tau$ calculated as
\begin{equation}
    \sigma_{\text{DCF}}(\tau) = \frac{1}{M-1} \sqrt{\sum_{ij} \left[ \text{UDCF}_{ij} - \text{DCF}(\tau) \right]^2}
\end{equation}
A positive correlation coefficient implies that the first time series leads the second, while a negative coefficient means it lags the second. 

\begin{figure*} 
    \centering  
    \includegraphics[width=0.4\textwidth]{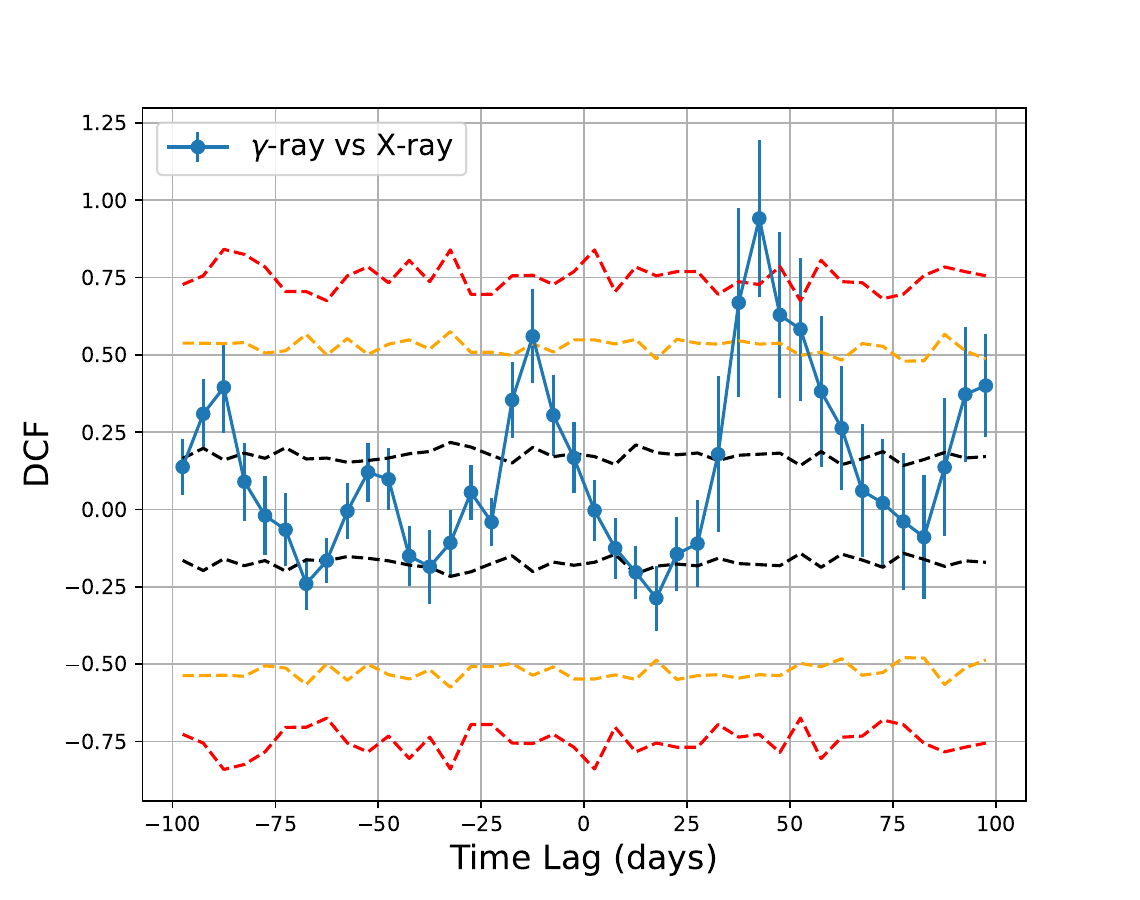}
    \includegraphics[width=0.4\textwidth]{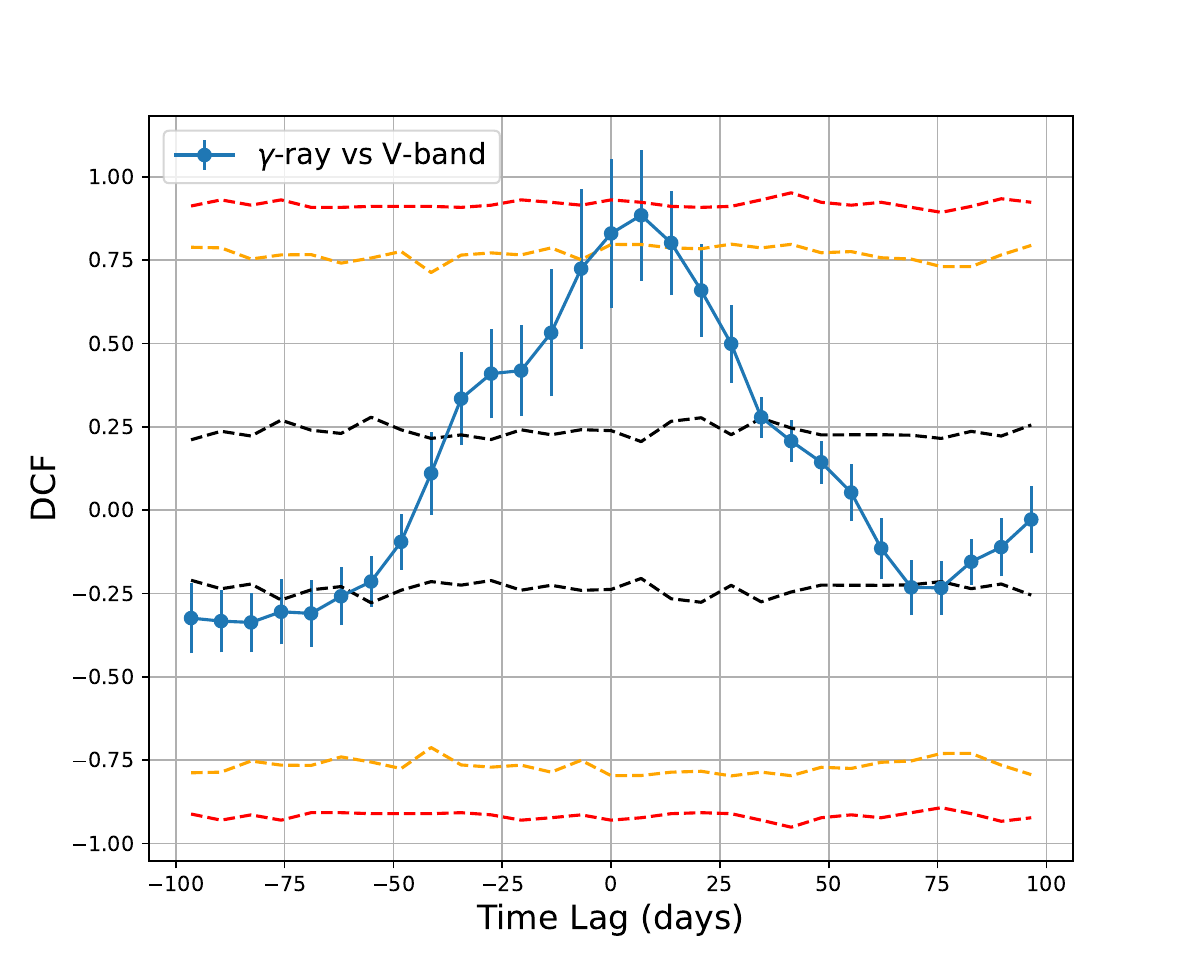}
    \includegraphics[width=0.4\textwidth]{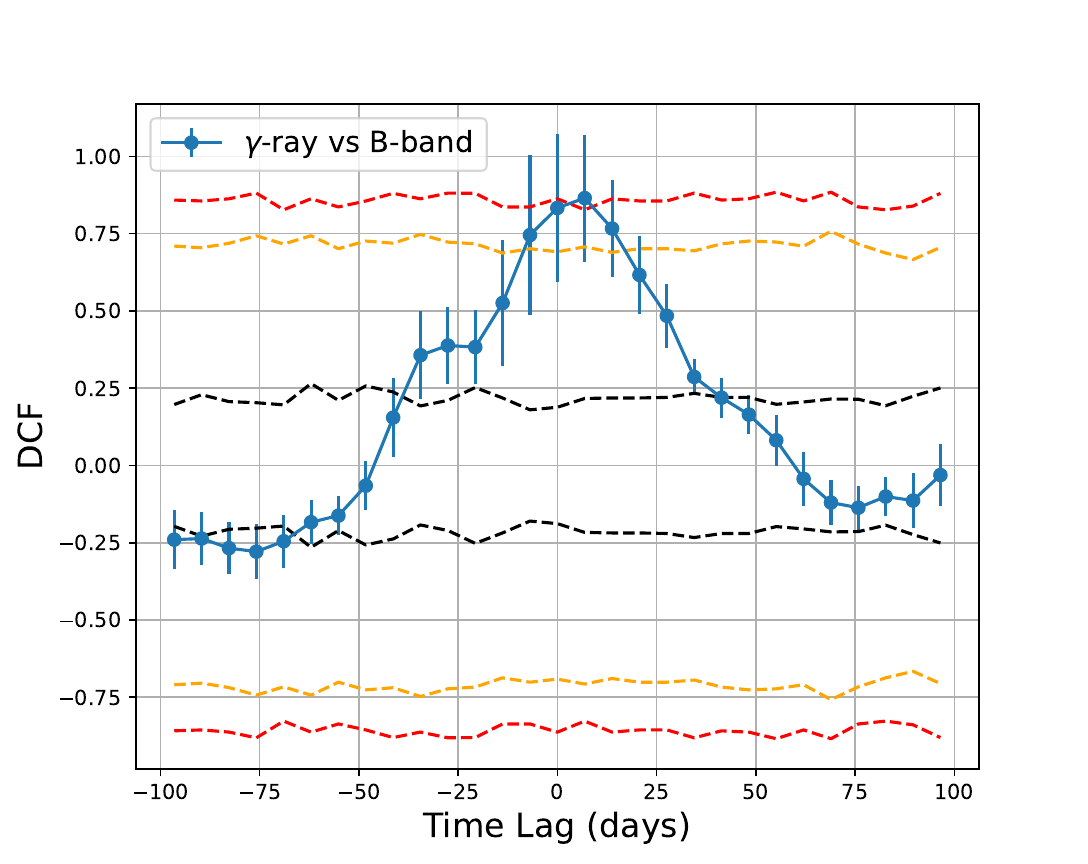}
    \includegraphics[width=0.4\textwidth]{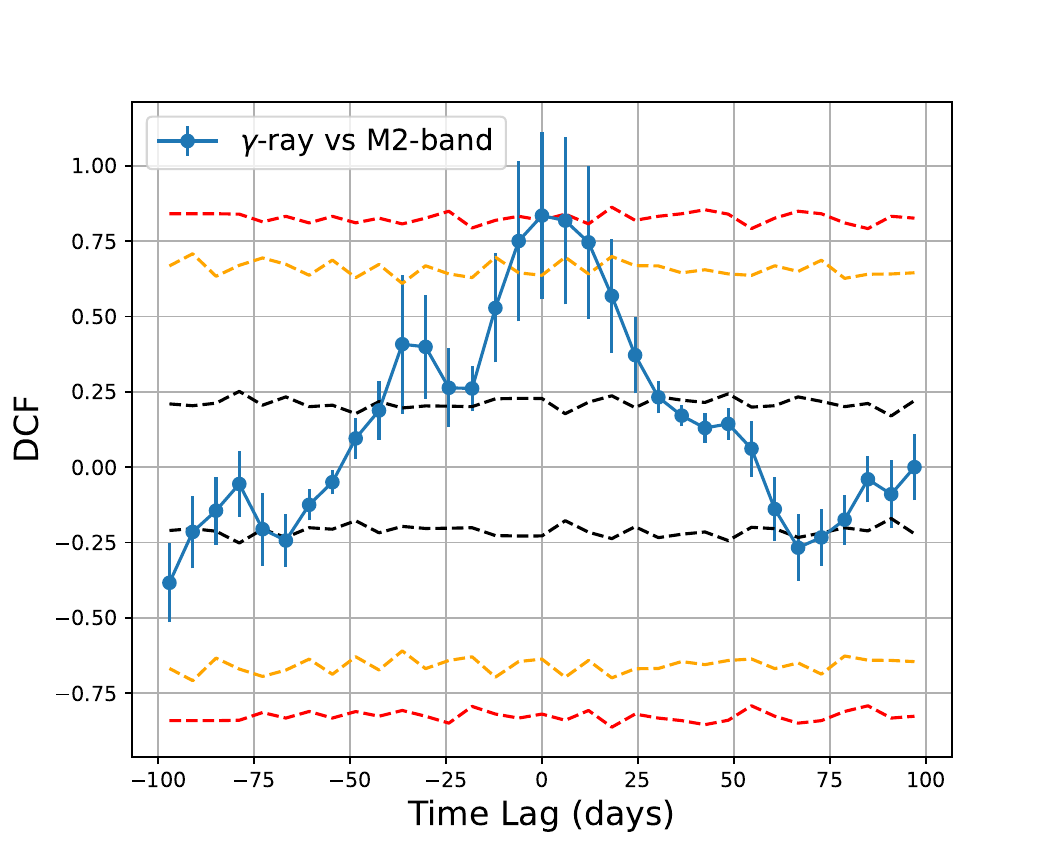}
    \caption{The outcomes of the discrete cross-correlation analysis conducted on the $\gamma$-ray with X-ray and optical (V, B) and UV (M2) band lightcurves during 59428 - 60312 MJD, repectively. The significance levels are represented by the black dashed line (68.27 per cent, 1$\sigma$ ), orange dashed line (95.45 per cent, 2$\sigma$ ), and red dashed line (99.73 per cent, 3$\sigma$).}
  \label{fig: corln_plt}
\end{figure*}

The  correlation study was carried for $\gamma$-ray with X-ray, optical (V, B, U), and UV (W1, M2, W2) wavebands, respectively. For robust correlation analysis, we chose time period between 59428 - 60312 MJD, due to availability of closely spaced data points in all bands. The selected period is marked by red vertical lines in Fig.\,\ref{fig: fvar}, and the results of correlation study are presented in Fig.\,\ref{fig: corln_plt}. We found that the $\gamma$-ray and X-ray flux points a show positive time lag of $42.5$ days with a high DCF value of $0.94\pm0.24$. This suggests that lightcurves of $\gamma$-ray and X-ray bands show a strong positive correlation ($>3\,\sigma$), with X-ray lagging behind the $\gamma$-ray emission. Moreover, correlation between $\gamma$-ray with all optical and UV (W1, W2) bands show a positive delay of $6.89$ days with a DCF values of $0.88\pm0.18$. Furthermore, it was found that the $\gamma$-ray and UV (M2) band show no significant lag, with a DCF value of $0.84\pm26$, suggesting a co-spatial emission region for these photons (see, Fig.\,\ref{fig: corln_plt}). We also performed the Z-transformed DCF (zDCF) analysis \citep{2013arXiv1302.1508A} to  verify weather the correlation coefficients were not due to DCF normalization. The zDCF yielded similar correlation coefficient values as the DCF results, with no new peaks detected. 

The significance of DCF is estimated by Monte Carlo (MC) method introduced by \cite{2014MNRAS.445..437M}. In the MC simulations, we generate 1000 random lightcurves for each band with a simple PL power spectral density (PSD) ($P(\nu)\propto \nu^{\beta}$) following the algorithm described in \cite{1995A&A...300..707T}. The spectral index $\beta$ for each band is determined by fitting the PSD of the actual observed lightcurve. Finally, the simulated lightcurve pairs are cross-correlated to determine the significance level at each time lag. All the combinations demonstrated a strong correlation around the 99\% level. 

\subsection{Spectral energy distribution and its modelling}
\label{sec:3.7}
In order to constrain the physical parameters responsible for simultaneous flux variations, we analyzed the broadband spectral properties of the source S5\,1044+71. The analysis was conducted by selecting different time intervals from the multi-wavelength lightcurve (see Fig.\,\ref{fig: multlc}).
Due to the pronounced short-term variability of the source, we employed Bayesian analysis to divide the 7-day binned $\gamma$-ray lightcurve into smaller segments. Each segment of the  $\gamma$-ray lightcurve is assumed to exhibit steady behavior in terms of the physical parameters. To select the Bayesian segments for broadband spectral analysis based on source activity and the availability of simultaneous $\gamma$-ray, X-ray, and UV/optical observations, we followed the approach mentioned in subsection~\ref{sec:3.1}. Here, the Bayesian block algorithm was adopted with a false alarm probability $p_0=0.05$ ($\geq 95\%$). The selected time segments (or states), with peak $\gamma$-ray flux above threshold line, are labeled as S1-A (56677 – 56756 MJD), S2-A (57539.5 – 57560 MJD), S2-B (57790 – 57841 MJD), S3-A (58827 – 58850 MJD) and S4-A (60019 – 60046 MJD). In addition to this, one quiescent flux state, characterized by a peak $\gamma$-ray flux below threshold line, is identified and labeled as the state QS (60102 – 60165 MJD). The selected time segments are marked by grey vertical strips as shown in  Fig.\,\ref{fig: multlc}. By comparing the results from different segments, one can identify the changes in the physical parameters responsible for the flux variations.

\begin{figure*} 
\label{fig:fsed}
    \centering
    \includegraphics[width=0.3\textwidth]{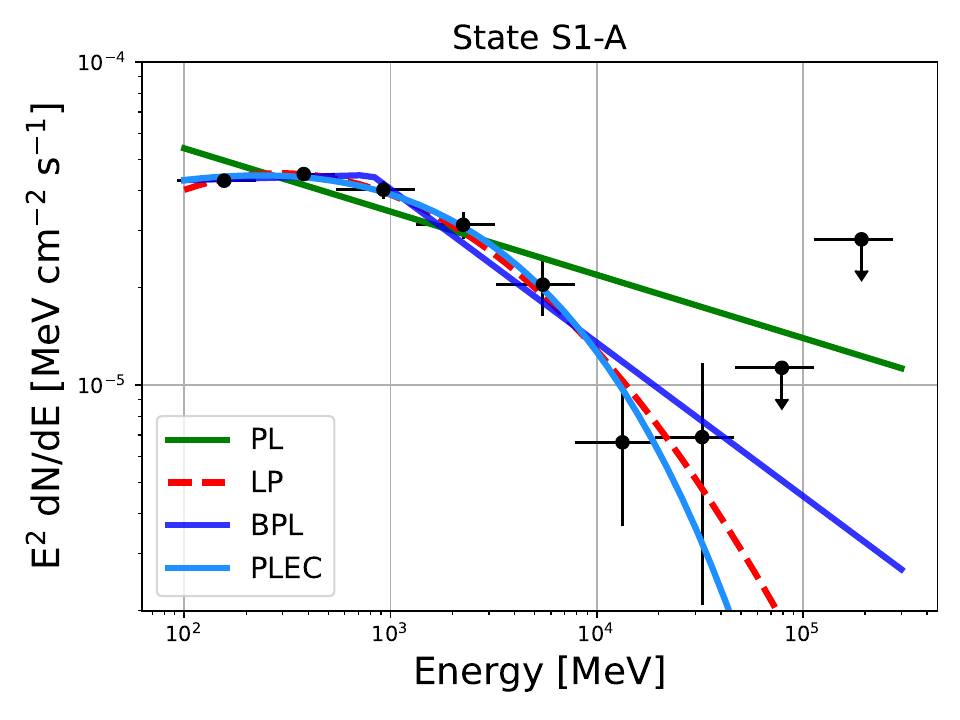}
    \includegraphics[width=0.3\textwidth]{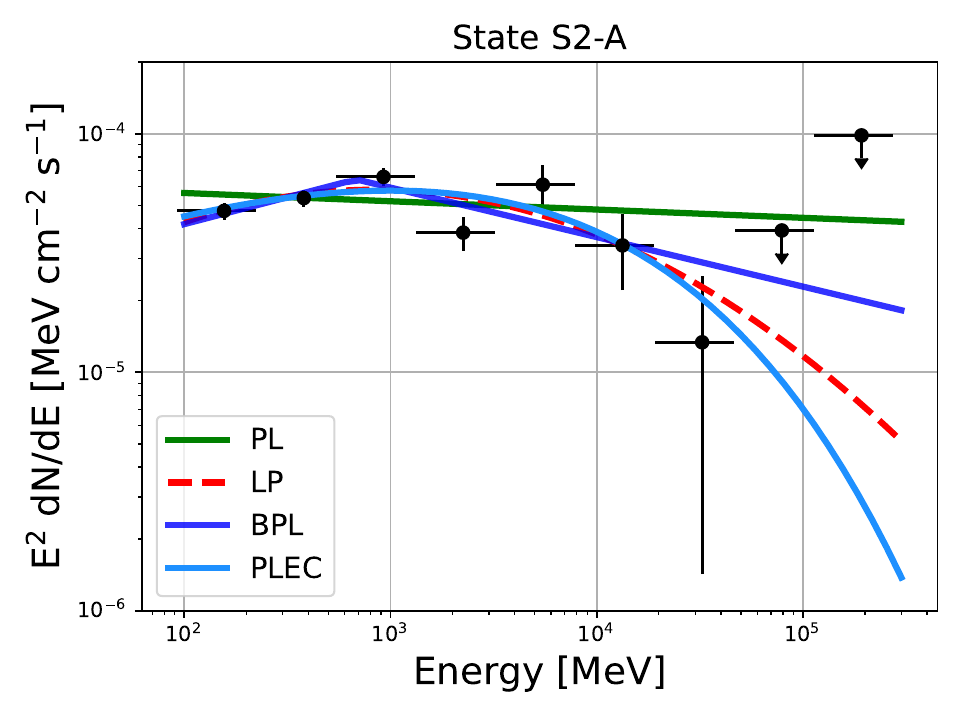}
    \includegraphics[width=0.3\textwidth]{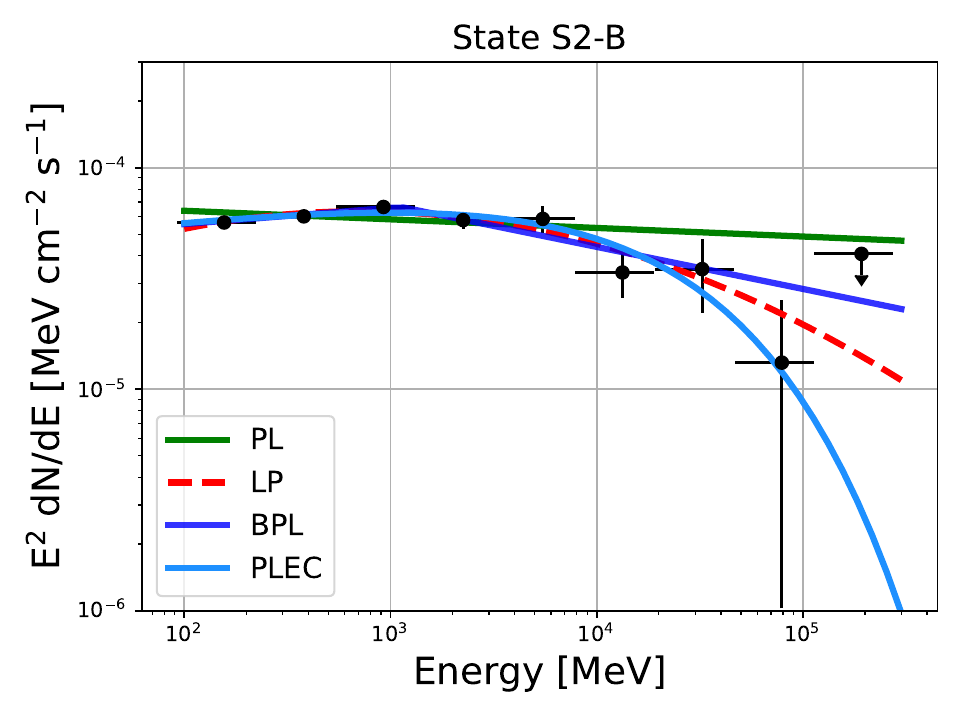}
    \includegraphics[width=0.3\textwidth]{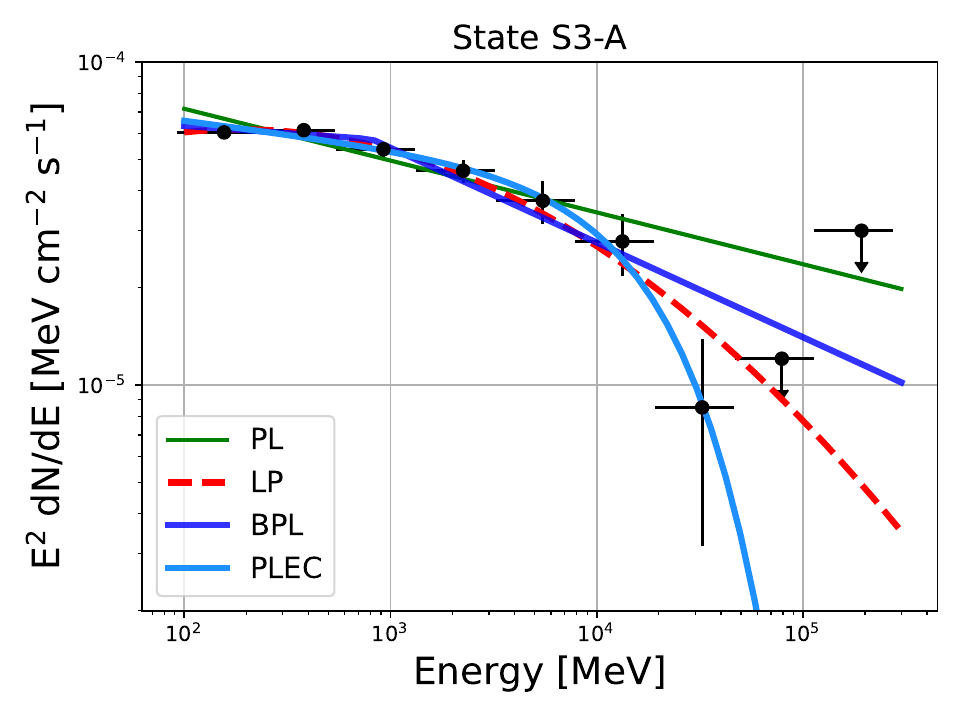}
    \includegraphics[width=0.3\textwidth]{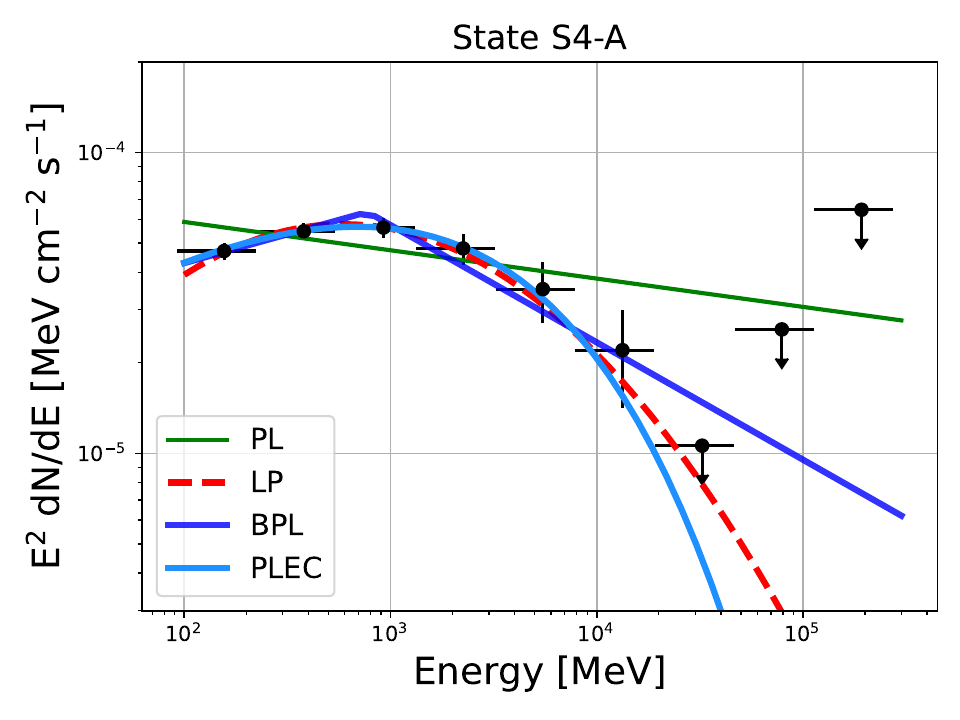}
    \includegraphics[width=0.3\textwidth]{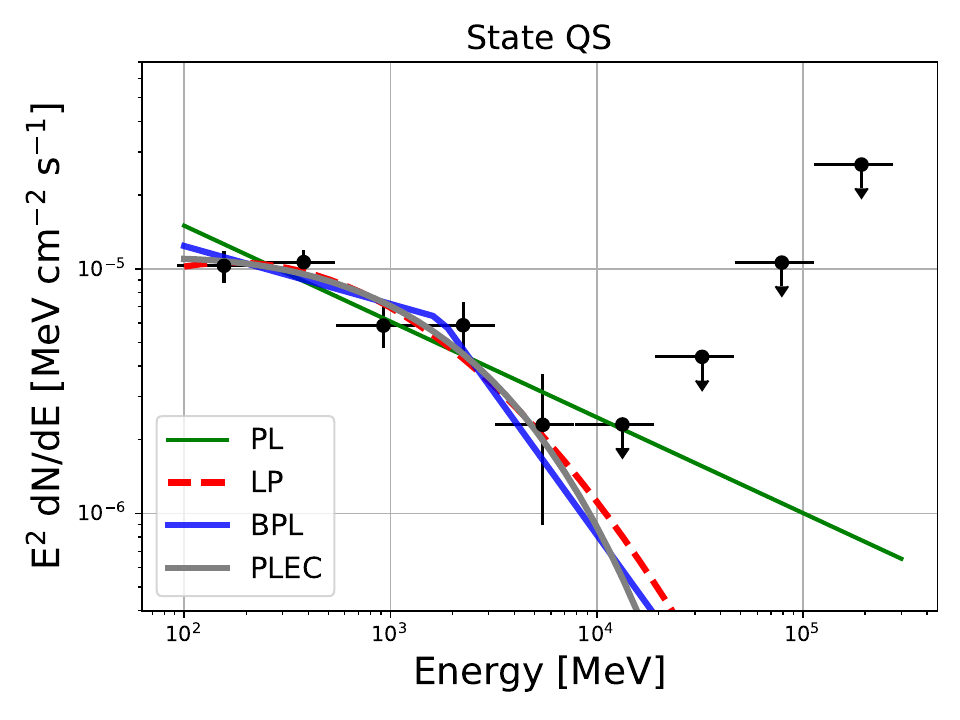}
    \caption{Plots of the $\gamma$-ray SEDs for different chosen epochs fitted with the spectral models shown. The details of the models are displayed within the insets in each plot.}
   \label{fig: fsed}  
\end{figure*}

\begin{figure*} 
    \centering
    \includegraphics[width=0.3\textwidth,angle=270]{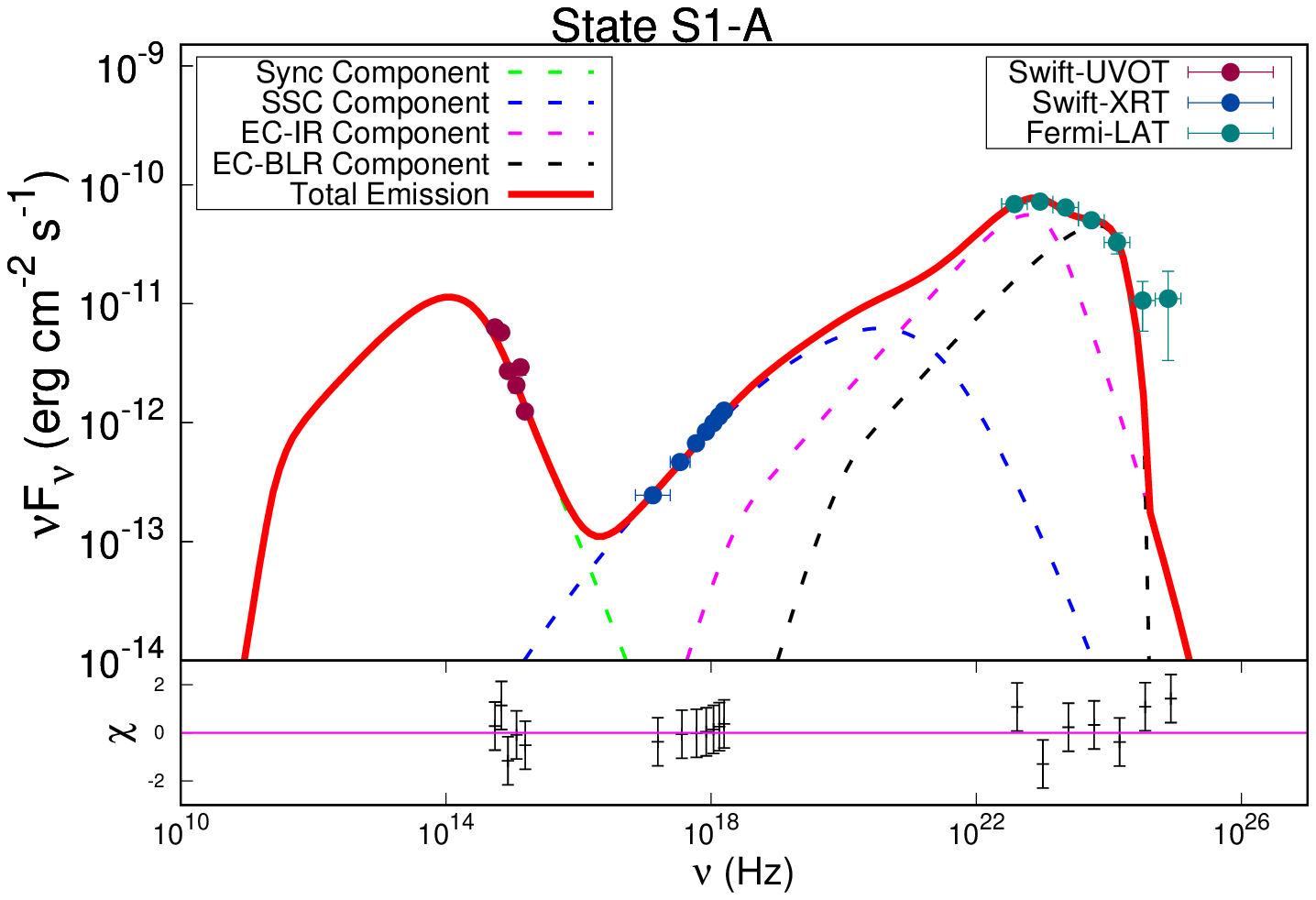}
    \includegraphics[width=0.3\textwidth,angle=270]{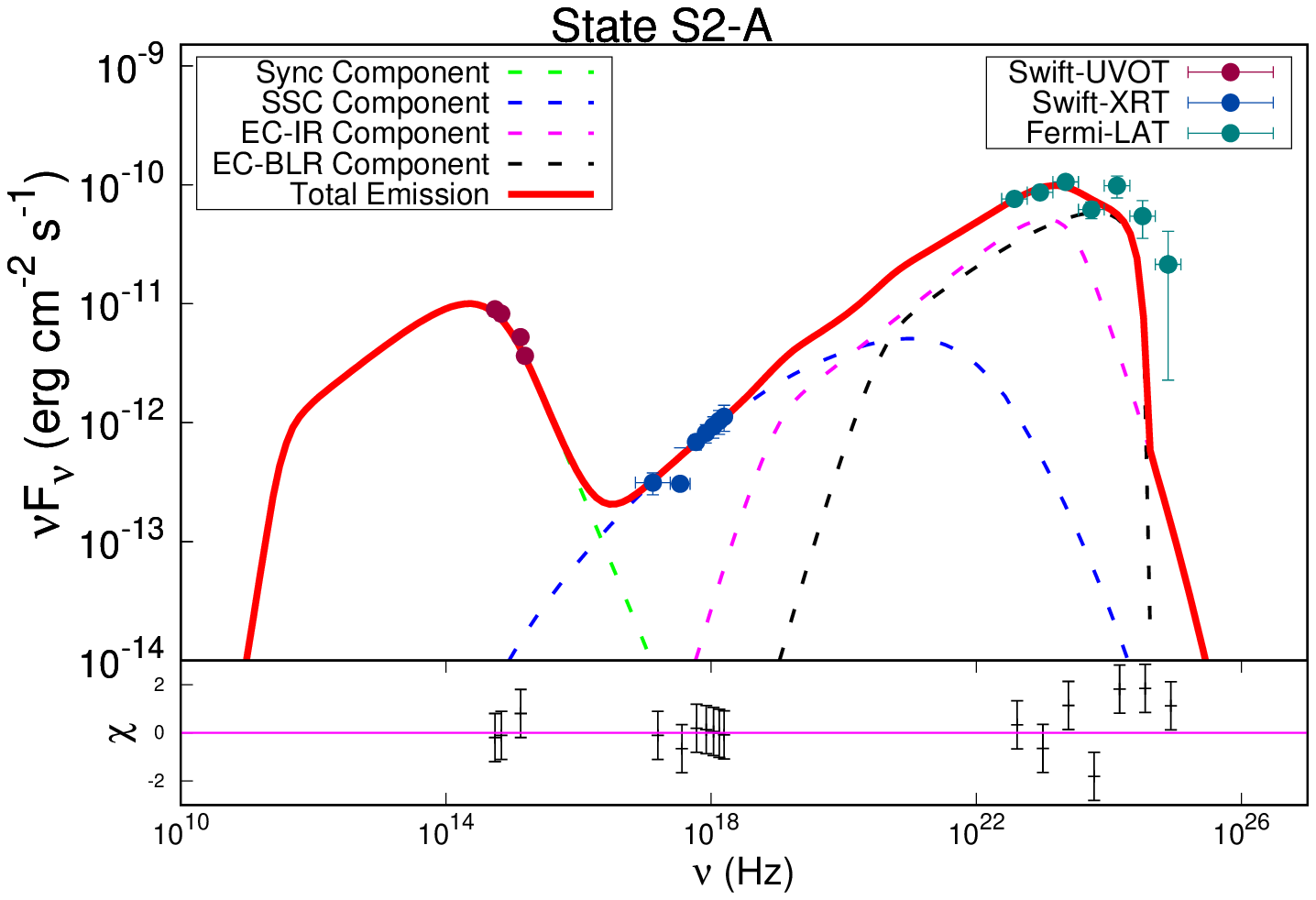}
    \includegraphics[width=0.3\textwidth,angle=270]{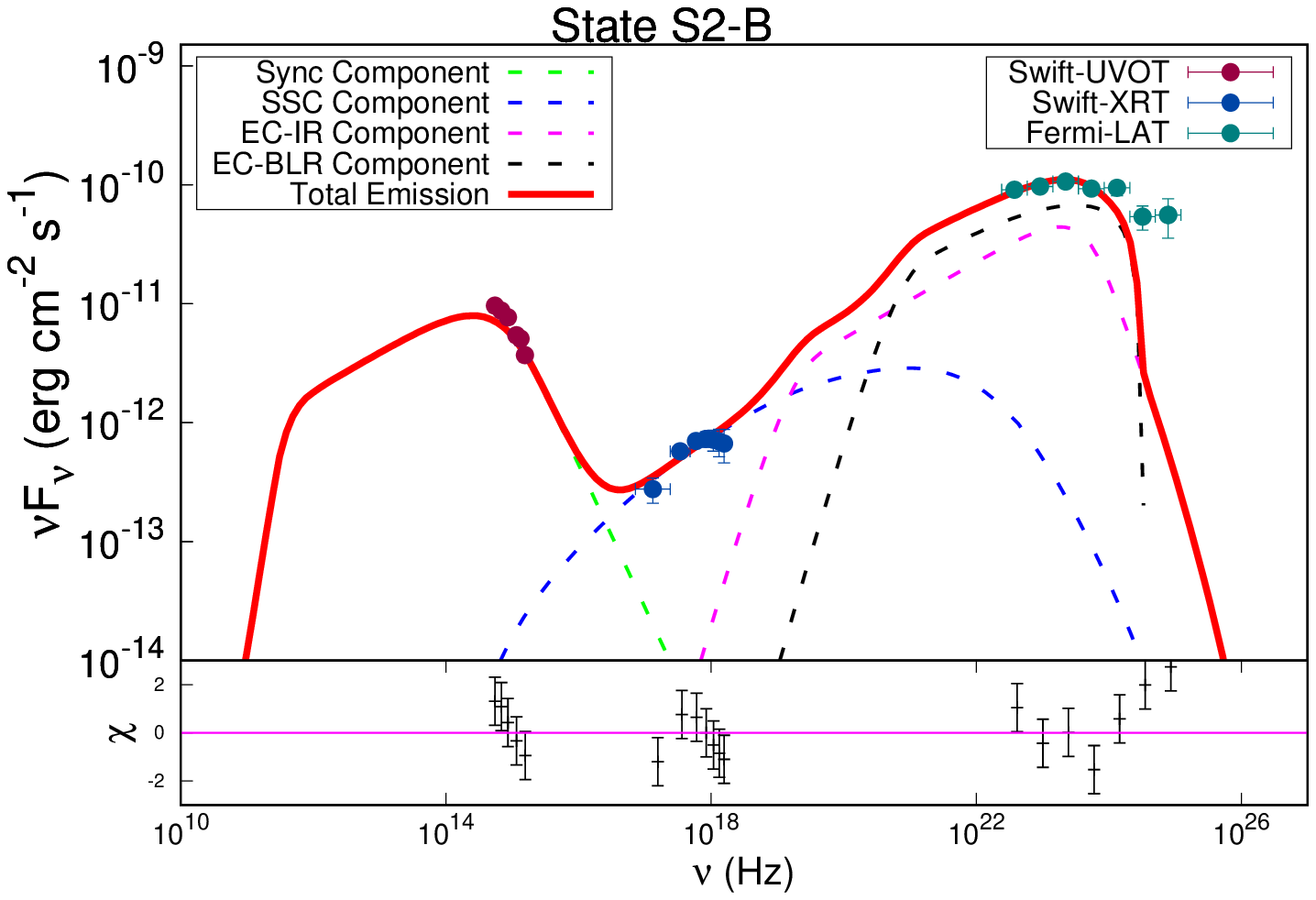}
    \includegraphics[width=0.3\textwidth,angle=270]{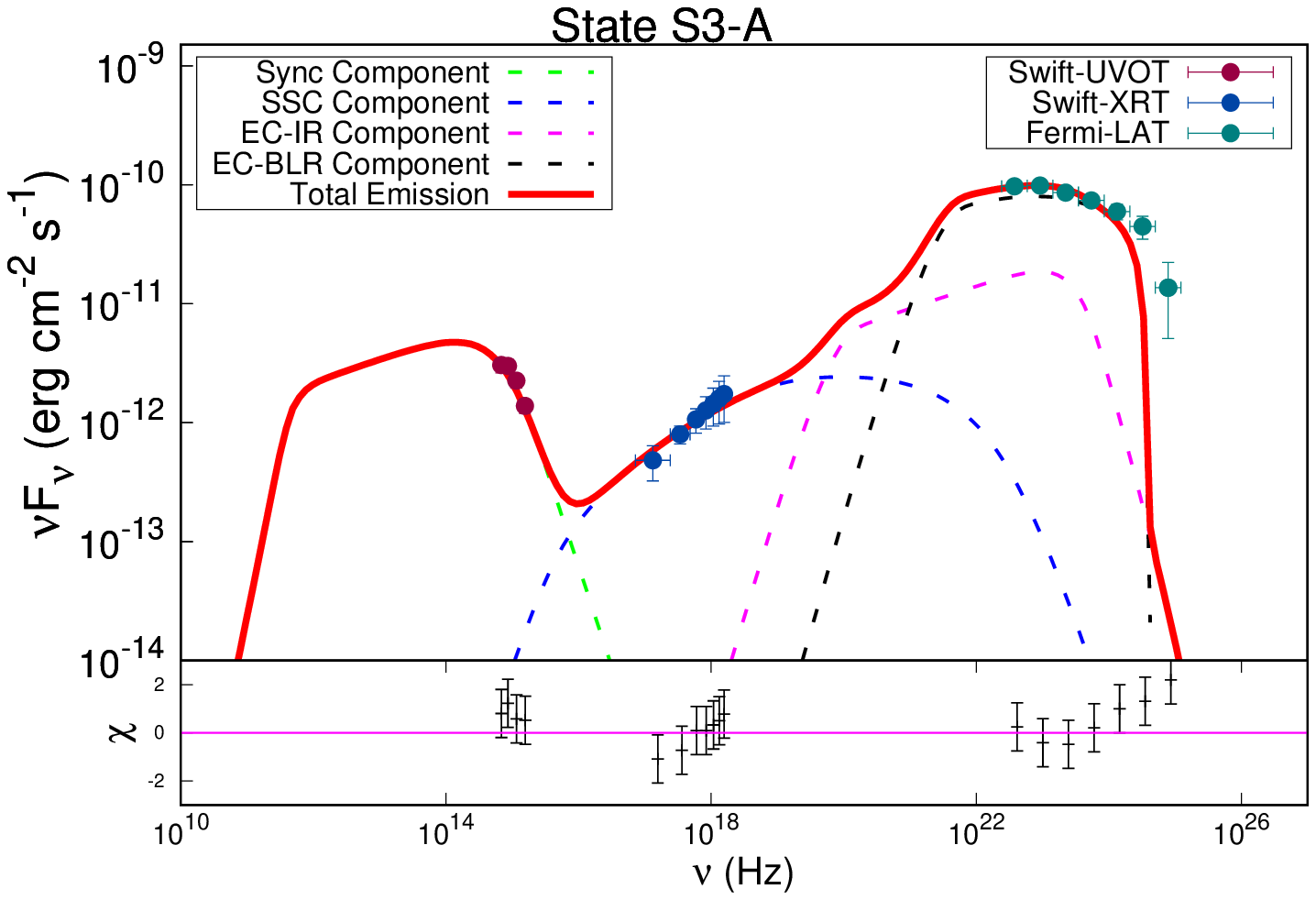}
    \includegraphics[width=0.3\textwidth,angle=270]{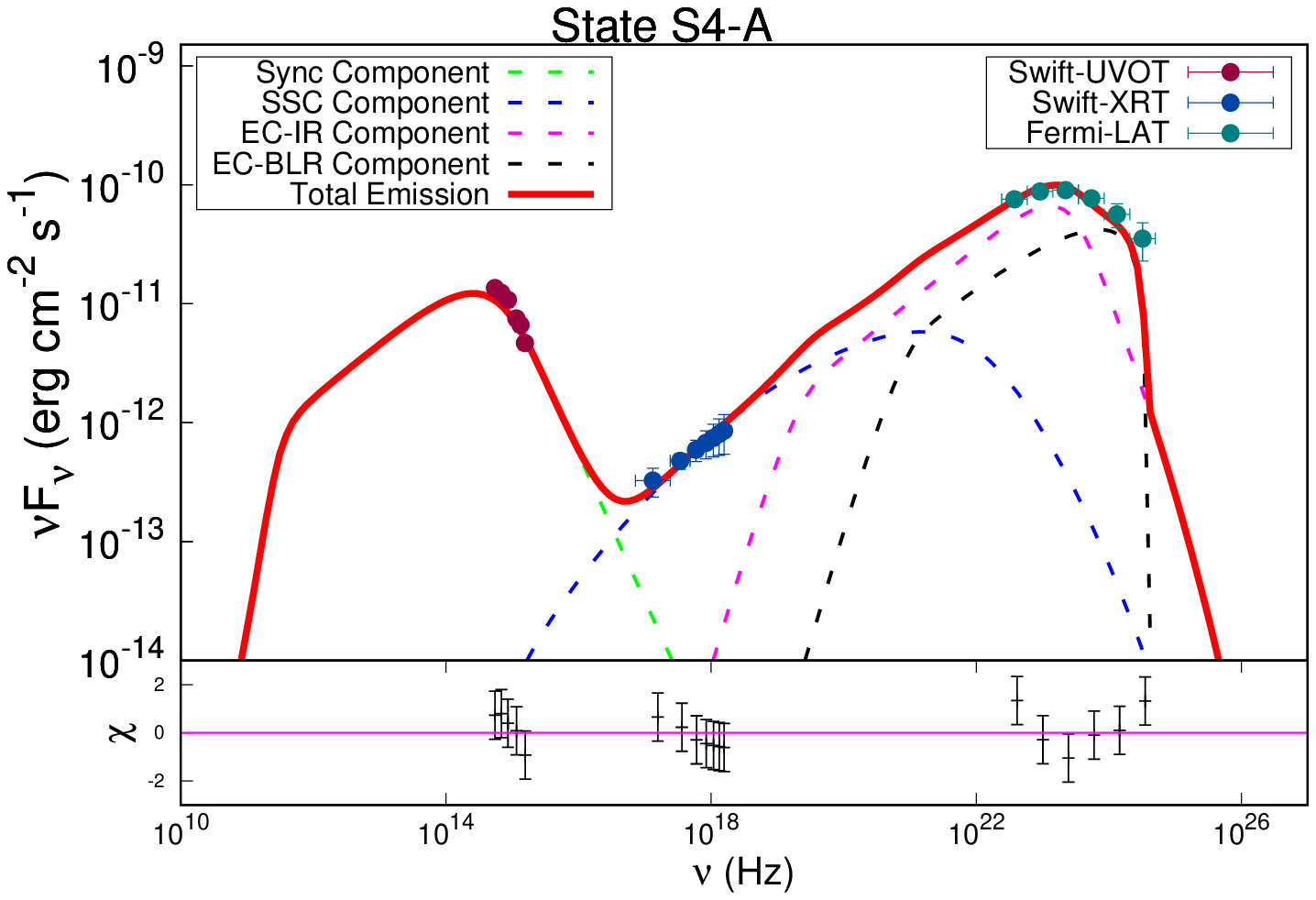}
    \includegraphics[width=0.3\textwidth,angle=270]{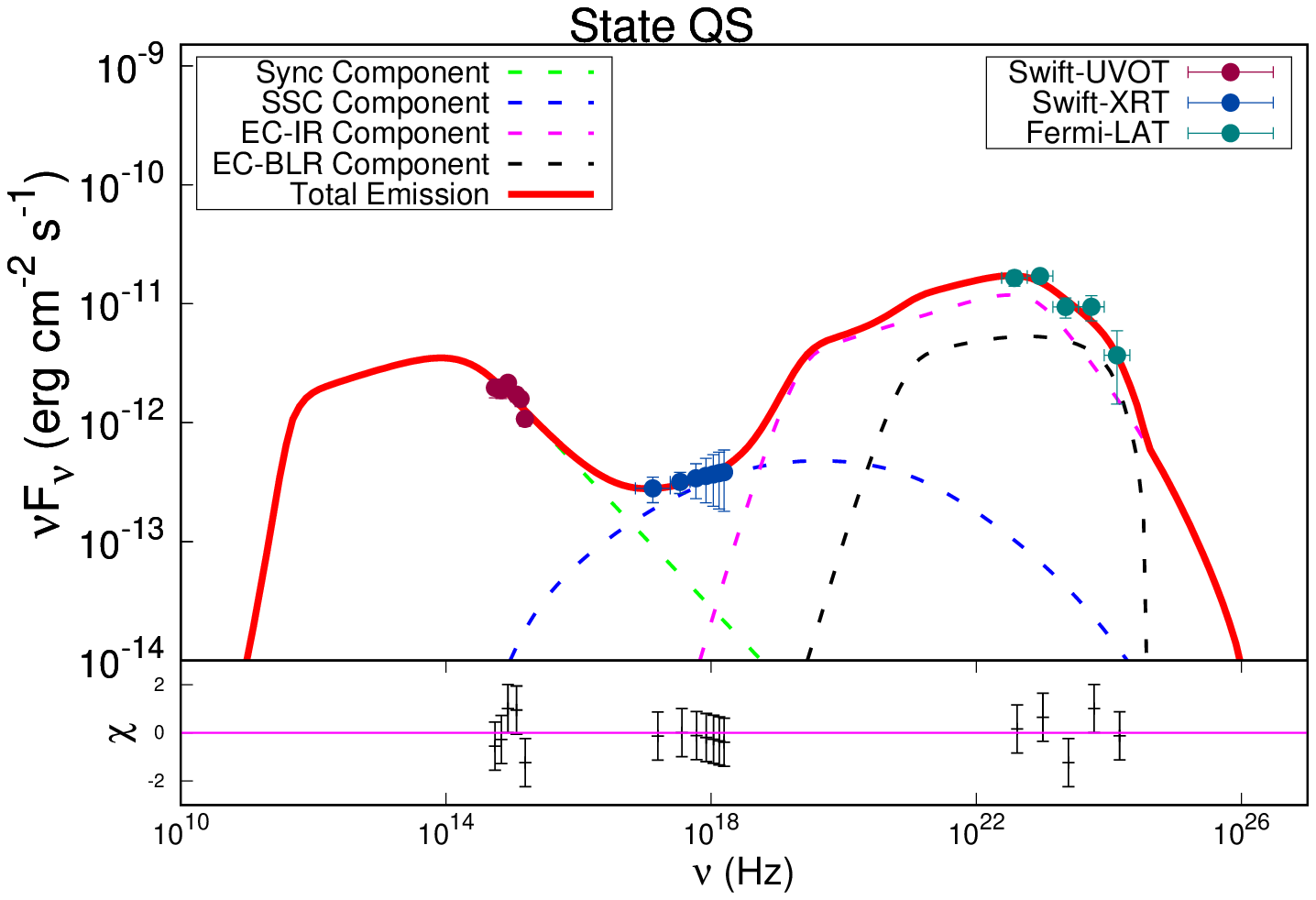}
    \caption{Broadband SEDs of S5\,1044+71 obtained during different flux states. The solid red curve represents the combined best-fitting synchrotron, SSC, and EC spectrum.}
    \label{fig: bsed}
\end{figure*}

The $\gamma$-ray spectral curvature in blazars can reveal the intrinsic distribution of emitting electrons and provide valuable information about the possible acceleration and cooling mechanisms within the jets. We performed the $\gamma$-ray spectral analysis of the source S5\,1044+71 by following the procedure outlined in subsection~\ref{sec:2.1}. The $\gamma$-ray spectra in the considered flux states were fitted by using log-parabola (LP), power-law (PL), broken power-law  and power-law with an exponential cutoff (PLEC) models. The statistical significance of curvature/break  in the $\gamma$-ray spectrum can be determined  by using the equation \citep{2012ApJS..199...31N}
\begin{equation}
\text{TS}_{\text{curve}} = 2 \left[ \log \mathcal{L}(\text{LP/BPL/PLEC}) - \log \mathcal{L}(\text{PL}) \right],
\end{equation}
where $\mathcal{L}$ is the likelihood function. The curvature/break in the spectrum is considered significant only if $\text{TS}_{\text{curve}} > 16$, corresponding to $4\sigma$ confidence level \citep{1996ApJ...461..396M}. The results of this analysis along with the obtained $\text{TS}_{\text{curve}}$ values are summarized in Table\,\ref{tab: fsed}, and the $\gamma$-ray spectra for the selected epochs are shown in Fig.\,\ref{fig: fsed}. The $\text{TS}_{\text{curve}}$ results indicate that $\gamma$-ray spectra of the states S1-A and S4-A exhibit a notable curvature with LP model providing the better fit statistics. However, the spectrum in these two states can equally well be described by the BPL/PLEC models in addition to the LP model. In case of state S3-A, the PLEC model provides a statistically significant fit to the $\gamma$-ray spectrum (see Fig.\,\ref{fig: fsed}). Furthermore, the $\gamma$-ray spectra for S2-A, S2-B and the quiescent state QS, are well described by the PL model. The $\gamma$-ray SED plots, along with the corresponding model fits, are presented in Fig.\,\ref{fig: fsed}.
Based on the results of the $\gamma$-ray spectral analysis, we chose the $\gamma$-ray SED points from the best-fitted model of each state for the broadband SED modelling. 
The X-ray and UV-optical spectral points in each state were acquired by following  the procedures mentioned in section~\ref{sec:2}. However, the X-ray spectra are well fitted by the model tbabs$*$pow, except for state S3-A, where the spectrum is best described by tbabs$*$logpar.
For the UVOT data, we note that its spectral shape deviates from the expected power-law model. This suggests that the optical/UV flux may include additional emission components beyond the jet, likely associated with the accretion disk \citep{2025JHEAp..4700400A}.
Since the uncertainties in the optical/UV flux measurements are relatively small, the broadband spectral fitting can be influenced by this energy band. To mitigate this bias and allow for a more representative fit using a simple power-law model, we added an additional systematic error of 5$\%$ to the optical/UV data.
The broadband SED points for states: S1-A, S2-A, S2-B, S3-A, S4-A and the quiescent state QS are shown in Fig.\,\ref{fig: bsed}.

\begin{table*}
	\centering   
 \caption{Results of $\gamma$-ray SED analysis for various observed states.}
\renewcommand{\arraystretch}{1.2} 
	\begin{tabular}{l l l l l l l l }
		\hline\hline
		Flare group/& $\text{F}_{0.1-300 \rm{GeV}}$& Power-Law & & & TS & $-\text{log}\mathcal{(L)}$ &$\text{TS}_{\text{curve}}$\\
		states& (10$^{-7}$ ph cm$^{-2}$ s$^{-1}$) &  $\Gamma$ & & & & &\\
		\hline
		S1-A       & 4.5$\pm$0.18   & -2.19$\pm$0.02 & -- & -- & 3112.13  & -21768.03 & --\\ 
		S2-A       & 5.4$\pm$0.36   & -2.03$\pm$0.02 & -- & -- & 1646.47 & -9243.03&--\\
		
		S2-B       & 6.1$\pm$0.23 & -2.03$\pm$0.02 & -- & -- & 4410.15 & -18591.85&--\\
		S3-A       & 6.1$\pm$0.22 & -2.15$\pm$0.02 & -- & -- & 4920.39  & -22664.31&--\\
		S4-A      & 5.3$\pm$0.25 & -2.71$\pm$0.03 & -- & -- & 2307.28 & -12554.29&--\\
	        QS     & 1.1$\pm$0.12 & -2.39$\pm$0.07 & -- & -- & 286.88 & -21244.86&--\\
		\hline
        \hline
		& & Log-Parabola & \\
		& & $\alpha$ & $\beta$ &  & & & \\
		\hline 
		S1-A     & 4.1$\pm$0.20 & 2.09$\pm$0.03 & 0.10$\pm$0.02 & -- & 3139.13 & -21754.99 &26.08\\
		S2-A     & 5.0$\pm$0.39 & 1.97$\pm$0.05 & 0.06$\pm$0.02 & -- & 1668.01  & -9239.10 &7.86\\
		S2-B     & 5.8$\pm$0.25 & 1.99$\pm$0.02 & 0.04$\pm$0.01 & -- & 4424.54 & -18586.24 &11.22\\
		S3-A     & 5.8$\pm$0.21 & 2.10$\pm$0.03 & 0.05$\pm$0.01 & -- & 4955.81 & -22657.55 &13.52\\
		S4-A   & 4.8$\pm$0.27 & 2.04$\pm$0.04 & 0.12$\pm$0.02 & -- & 2357.31& -12541.28 &26.02\\ 
		QS     & 0.95$\pm$0.14 & 2.38$\pm$0.10 & 0.13$\pm$0.07 & -- & 289.89 & -21242.71 &4.30\\
		\hline
        \hline
		& & Broken Power-Law & & $\epsilon_{\rm break}$& &  \\
		& & $\Gamma_1$ & $\Gamma_2$ & (GeV) & & &\\
		\hline
		S1-A       & 4.1$\pm$0.20 & -1.98$\pm$0.06 & -2.41$\pm$0.09 & 0.88$\pm$0.23 & 3137.37 & -21755.69 &24.68\\
		S2-A       & 4.9$\pm$0.41 & -1.77$\pm$0.16 & -2.20$\pm$0.09 & 0.67$\pm$0.39 & 1676.57  & -9238.62 &8.82\\
		S2-B      & 5.8$\pm$0.26 & -1.92$\pm$0.09 & -2.13$\pm$0.13 & 1.12$\pm$0.41 & 4428.42 & -18586.87& 9.96\\
		S3-A       & 5.9$\pm$0.23 & -2.04$\pm$0.04 & -2.29$\pm$0.05 & 0.84$\pm$0.09 & 4957.94 & 22659.47 &9.68\\
		S4-A      & 4.8$\pm$0.27& -1.80$\pm$0.08 & -2.38$\pm$0.10 & 0.78$\pm$0.22 & 2353.42& -12542.58 &23.42\\
                QS            & 0.98$\pm$0.24 & -2.23$\pm$0.40 & -3.16$\pm$4.51 & 1.74$\pm$9.61 & 290.19 & -21242.57 &4.58\\
	\hline
    \hline
		& & PLExpCutoff & &$\epsilon_{\rm cutoff}$  & & \\
		& & $\Gamma_{\rm PLEC}$ &
                &(GeV)  && &\\
	
        \hline
		S1-A    & 4.2$\pm$0.6 & -1.71$\pm$0.06 &--
                   &0.5$\pm$0.10  & 3139.4 & -21755.30 &25.46\\
		S2-A    & 5.08$\pm$0.03 & -1.73$\pm$0.14 &--
                   & 3.01$\pm$0.54  & 1663.94  & -9239.06 &7.94\\
		S2-B    & 5.8$\pm$1.21  & -1.89$\pm$0.12
                    & --
                    &1.8$\pm$0.25 & 4423.29 & -18585.24 &13.22\\
		S3-A    & 5.9$\pm$0.23  & -2.08$\pm$0.07 &-- 
                    &2.1$\pm$0.97 & 4947.89 & -22656.04 &16.54\\
	        S4-A     &4.9$\pm$1.18  &-1.53$\pm$0.08 &-- 
                   & 0.5$\pm$0.17 & 2353.80& -12541.13 &26.32\\  
                QS     & 0.96$\pm$0.13 & -1.80$\pm$0.30 &--
                   &0.50$\pm$0.95 & 290.30 & -21242.59 &0.24\\
	\hline
	\hline
	\end{tabular}%
 \label{tab: fsed}
\end{table*}

The broadband spectral energy distribution (SED) of the source, in the selected flux states, are modelled using a one-zone leptonic model considering synchrotron, SSC and EC processes \citep{2017MNRAS.470.3283S,2018RAA....18...35S,2025JHEAp..4700400A}. This model considers the emission region as a spherical blob with radius R, moving along the blazar jet with a bulk Lorentz factor \( \Gamma \) and oriented at an angle \( \theta \) to the observer. The emission region contains a non-thermal electron population characterized by a broken power-law distribution
\begin{align} 
\label{eq:broken}
	N(\gamma)\,d\gamma = \begin{cases}
		K\,\gamma^{-p}\,d\gamma&\textrm{for}\, \mbox {~$\gamma_{\rm min}<\gamma<\gamma_b$~} \\
		K\,\gamma_b^{q-p}\gamma^{-q}\,d\gamma&\textrm{for}\, \mbox {~$\gamma_b<\gamma<\gamma_{\rm max}$~}
		\end{cases},
\end{align}
where, K is the normalization factor,  \( \gamma \) represents the dimensionless energy of the electrons; \( \gamma_{\text{min}} \) and \( \gamma_{\text{max}} \) are the minimum and maximum dimensionless electron energies, respectively; \( p \) and \( q \) are the low- and high-energy spectral indices of the distribution; and \( \gamma_b \) is the break energy. 
These relativistic electrons, in the presence of a tangled magnetic field \( B \), produce synchrotron emission. Additionally, they also undergo IC scattering which includes the SSC and EC processes. For the EC emission, the target photons are assumed to be dominated by Ly-\(\alpha\) photons from the BLR and thermal IR photons from the molecular torus.
The emissivity functions for the radiative processes (synchrotron, SSC and EC) are computed numerically, and the resulting routines are incorporated as a local model in the \texttt{XSPEC}. For numerical simplicity, the BLR emission is approximated as a blackbody at a temperature of  42,000\,K, equivalent to the Ly-$\alpha$ line ($2.47\times 10^{15}$\,Hz) while, the emission from the molecular torus is treated as a blackbody with a temperature of 1000\,K. 
Under the synchrotron, SSC, and EC emission processes, the observed spectrum is primarily governed by ten parameters. Among these, \( K \), \( p \), \( q \), and \( \gamma_b \) describe the electron energy distribution; \( B \), \( R \), and \( \theta \) characterize the macroscopic properties of the emission region; and two additional parameters represent the frequency and energy density of the external photon field.
However, due to limited observational constraints in different energy bands, we adopted a minimalistic emission model. In this model, the spectral fit for different selected states was performed by setting all parameters free, except $\theta =2^\circ$, $\text{R}=1.82\times10^{16}$ cm, $\gamma_{max}=2\times10^{8}$ and $f$ (fraction of external scattered photons). The values of $f$ for IR and BLR photons were fixed at $\sim$ $5.31\times10^{-2}$ and $\sim$ $10^{-7}$, respectively.
Unlike other blazars where a single external photon field (either from the BLR or the IR torus) often provides a satisfactory fit to the \(\gamma\)-ray spectrum, our source required contribution from both the BLR and the IR torus to reproduce the observed spectra in the selected flux states. The best-fit spectral model, which includes synchrotron, SSC, and EC (IR + BLR) components, along with the observed SED points for all selected states, is shown in Fig.\,\ref{fig: bsed}, and the corresponding fitting parameters are provided in Table\,\ref{tab: bsed}.

\begin{table*}
	\centering
\caption{Optimal parameters of the broadband SED modeling during different selected periods.}
	
	\begin{tabular}{llllllll}
		\hline\hline
Parameter& Symbol & S1-A & S2-A &S2-B&S3-A&S4-A&QS \\
		\hline
Low energy Particle index    &  $p $     &    $1.73_{-0.19}^{+0.21}$ &    $2.11_{-0.21}^{+0.22}$  & $2.27_{-0.22}^{+0.23}$&$2.61_{-0.20}^{+0.24}$ &$2.07_{-0.17}^{+0.18}$ &$2.64_{-0.28}^{+0.26}$\\ 
High energy Particle index    &     $q$    &    $5.80_{-0.71}^{+0.80} $  &$    5.70_{-0.68}^{+0.70}$ & $5.06_{-0.69}^{--}$&$6.36_{-0.70}^{--}$&$5.45_{-0.61}^{+0.61}$&$4.12_{-0.66}^{+0.61}$ \\
Break Lorentz factor    &     $\gamma_b$  &   $2092_{-408}^{--}$ &   $2871_{-431}^{+466}$ & $3588_{-461}^{+513}$ &$3377_{-481}^{+496}$&$3009_{-445}^{+479}$ &$1622_{-321}^{--}$\\ 
Minimum electron Lorentz factor&$\gamma_{min}$& $11.31_{-2.11}^{+2.23}$ & $18.48_{-3.12}^{--}$ & $51.37_{-4.65}^{--}$ &$50.01_{-4.31}^{--}$&$27.88_{-3.62}^{+3.81}$ &$26.6_{-4.03}^{+4.26}$\\
Magnetic Field (G)  & $B$ & $1.21_{-0.22}^{--}$ & $1.24_{-0.15}^{+0.18}$ & $0.96_{-0.23}^{--}$ &$0.98_{-0.17}^{+0.22}$&$1.20_{-0.11}^{+0.15}$ &$1.81_{-0.19}^{+0.25}$\\  
Total jet power (log) & P$_{\rm jet}$ & $45.86$ & $45.94$ & $45.67$&$45.92$ &$45.72$ &$45.90$\\

 Equipartition& $U_e/U_B$  & $1.14$ & $1.22$ & $1.94$ & $2.83$ &$0.88$ &$0.42$  \\

Reduced-$\chi^2$ ($\chi^2/dof$)   &  $\chi^2_{red}$      &    $13.25/12$ &    $14.88/12$  & $21.13/13$&$16.50/11$ &$9.62/11$&$7.36/10$\\ 

  \hline\hline
        
\end{tabular}%
\label{tab: bsed}    
\end{table*}

\section{SUMMARY AND DISCUSSION}
\label{sec:4}
We analyzed 16 years of \textit{Fermi}-LAT observations to probe the high-energy emission characteristics of the source S5\,1044+71. The source displayed both long-term and short-term variability. During this period, the 3-day binned $\gamma$-ray lightcurve showed multiple outburst phases with a peak flux of $(1.1 \pm 0.2) \times 10^{-6} \, \text{ph} \, \text{cm}^{-2} \, \text{s}^{-1}$ observed at 57751 MJD. This represents the highest 3-day binned $\gamma$-ray flux ever detected from the source. The 3-day binned $\gamma$-ray lightcurve is also characterized by the temporal structure of the flare profiles.
 During our analysis, eight peaks exhibit symmetric profiles with $T_r \sim T_d (-0.3 \leq \zeta\leq 0.3)$, while two peaks show moderate asymmetry, with one peak showing $T_r < T_d (\zeta<-0.3)$ and other $T_r > T_d (\zeta>0.3)$ (see Table\,\ref{tab: rt_dt}). 
 Hence, most of the peaks have symmetric temporal profiles within the error bars. 
 The detection of symmetric or asymmetric flares is closely related to the physical characteristics of the emitting region, such as spectral properties, the location of the peak flux with respect to energy, and the relevant acceleration and cooling timescales. In the case of a relativistic plasma propagating through a shock in the jet, the rise time corresponds to the duration over which the plasma enters the shock region, resulting in an increase in the number of accelerated electrons and, consequently, a rise in flux. The decay time reflects the period during which the plasma leaves the shock region, leading to a steady decrease in the number of energetic electrons and hence a decline in flux. Symmetric flares, where the rise and decay times are comparable, are expected when the radiative cooling time of the emitting particles is much shorter than the timescale of the shock–plasma interaction \citep{1979ApJ...232...34B,2012ApJ...749..191C}. In this scenario, the observed symmetry in the temporal profiles suggests that both rise and decay phases are primarily governed by the plasma crossing time through the shock front. Conversely, asymmetric flares may occur when the interaction time is shorter than the characteristic acceleration or cooling timescales. Furthermore, \citep{2019MNRAS.484.3168S} have shown that such asymmetry cannot be attributed merely to variations in the efficiency of the acceleration process, as such changes would produce similar behavior in both the rising and decaying phases. Instead, they demonstrated that asymmetry can arise due to a shift in the SED peak energy during the flare.
For the source S5\,1044+71, the symmetric temporal profiles of the flares indicate that the rise and decay timescales are primarily governed by the crossing time of the radiation or a disturbance from a dense plasma blob moving through the shock front in the jet region. 
In a flare where the acceleration timescale is much shorter than the cooling timescale, the particles are energized more rapidly than they can lose energy through radiative processes such as synchrotron and inverse Compton cooling. Consequently, the acceleration region moves almost all the particles to their maximum energies. Once this acceleration phase is complete, the particles begin to cool slowly. As a result, the rise and decay times are not similar, leading to an asymmetric temporal profile \citep{1999APh....11...45K}. In this work, an asymmetric temporal proﬁle would be expected to arise from the fast injection of accelerated
particles or an escape from the emission region \citep[see, e.g.,][]{2001ApJ...554....1S,2011ApJ...727...21J,2019MNRAS.482..743R}

The variability timescale of the source can be used to
constrain the size of the emission region. An upper limit on the physical size of the $\gamma$-ray emission region can be calculated as
\begin{equation}
\centering
R \leq c \, t_{\text{var}} \, \delta \, /(1 + z).
\end{equation}
Using  the Doppler factor of $\delta=20$ obtained through broadband SED modelling and a variability timescale of $t_\text{{var}}=4.5$ hr, we derive an upper limit on the size of the $\gamma$-ray emission region radius R $\sim$ $2.089\times10^{16}$\,cm. This result is consistent with the results obtained by SED modelling but smaller than the value estimated by \cite{2024ApJ...962...22Z}. The upper limit on the angular
size [microarcseconds ($\mu as$)] of the emission region can be calculated by \citep{2013A&A...552A..11R,2013A&A...557A..71R} as:
\begin{equation}
\phi \leq 173 \,\frac{ t_{\text{var}}} {d_L}\, \delta \, (1 + z) \, \text{$\mu as$},
\end{equation}
where $t_{\text{var}}$ is the variability timescale in years and $d_L=7.9$ Gpc is the luminosity distance. We find that the angular size of the $\gamma$-ray emission region is $\sim 0.49 \mu as$. The distance of the emission region along the jet axis can be constrained using the observed variability timescale, \( d_\gamma \sim 2 c t_{\text{var}} \Gamma^2/({1 + z}) \). For blazars, the bulk Lorentz factor (\( \Gamma \)) of the radiating blob is typically \( \Gamma \sim \delta \) (Doppler factor). Based on this, we calculate \( d_\gamma \sim 0.0292\,\text{pc} \) ($9.015\times10^{16}\,$cm) which places the emission region near the BLR ($\text{R}_\text{{BLR}} \sim 3.4\times10^{17}$cm) \citep{2024ApJ...962...22Z}. Hence, our result also indicates that a single-zone model, including external Compton emission from both the IR and the BLR components, should be able to accurately fit the broadband SEDs of the source.

The source displayed three major $\gamma$-ray outburst phases, each with a peak flux exceeding the threshold line (see, Fig.\,\ref{fig: ph_ts}). The 3-day binned flux distribution of the flare epochs S1, S2, and S4 is consistent with a log-normal distribution. However, the binned flux distribution during the flare epoch S3 and the entire 3-day binned $\gamma$-ray lightcurve exhibit  double log-normal behavior.
This indicates that, unlike the other flare epochs, S3 may involve either multiple emission components or rapidly evolving physical conditions within the emission region. These features have also been reported in other bright blazars \citep[see, e.g.,][]{2016ApJ...822L..13K,2018RAA....18..141S,2020A&A...634A.120A,2020MNRAS.491.1934K}. Hence, our results support the notion that the $\gamma$-ray emission regions or the dominant variability mechanisms can vary across different flare epochs, consistent with the conclusions of \cite{2017ApJ...844...62P}.
As demonstrated  by \citet{2018MNRAS.480L.116S}, that linear Gaussian fluctuations in the particle acceleration timescale naturally produce a single log-normal flux distribution, while variations in particle escape timescales can lead to non-Gaussian flux distribution. In this context, the single log-normal flux distributions observed in S1, S2, and S4 are consistent with a scenario dominated by Gaussian fluctuations in acceleration processes. In contrast, the double log-normal distribution observed in S3 implies either the coexistence of two emission components or a regime where variations in escape timescales become significant, leading to double log-normal component in the flux distribution.

Over the past 16 years, the source has demonstrated diverse flux variability, including periods of correlated GeV, X-ray, and optical/UV variations, as well as $\gamma$-ray flares without corresponding X-ray counterparts. We utilized the advanced DCF to investigate potential correlations between the $\gamma$-ray, X-ray and optical/UV lightcurves. Our analysis revealed a strong positive correlation between the variations in optical/UV and $\gamma$-ray fluxes, particularly during the period 59428–60312 MJD, where $\gamma$-ray emission was found to lead the optical/UV emission by few days. During the cross-correlation study of bright blazars, \cite{2019ApJ...880...32L} found that the time lag between the optical and $\gamma$-ray bands is generally small. Similarly, \cite{2023MNRAS.519.6349D} reported that the time lag between these bands is generally consistent with zero. These findings suggest that the optical and $\gamma$-ray flux emission regions are co-spatial, supporting the leptonic emission model for blazars, with same population of relativistic electrons responsible for both optical and $\gamma$-ray emissions. Since both fluxes are proportional to the number of emitting electrons, it is plausible that variations in the electron population within the jet drive the observed correlation between these flux variations, as suggested by \cite{2011A&A...529A.145D}. Furthermore, the $\gamma$-ray and X-ray flux variations exhibited a strong positive correlation ($>$ 3$\sigma$) with X-ray lagging behind the $\gamma$-ray emission by 42.5 days. Additionally, analysis using zDCF also revealed a positive correlation between the emissions across different bands.
The DCF between the $\gamma$-ray and X-ray bands does not exhibit a single dominant peak but instead shows two peaks at different time lags (see, Fig.\,\ref{fig: corln_plt}), indicating a complex correlation between the two bands. 
One possible explanation is the involvement of multiple external seed photon fields in the inverse Compton process responsible for the $\gamma$-ray emission, such as infrared photons from the dusty torus and photons from the broad-line region. In contrast, the X-ray emission is likely dominated by the SSC process, in which the same electron population produces synchrotron photons and subsequently upscatters them to X-ray energies. This difference in emission mechanisms can naturally lead to a more complex correlation structure between the $\gamma$-ray and X-ray bands.
Consistent with this picture, the flux–flux correlation analysis indicates a strong correlation between the $\gamma$-ray and optical/UV bands, whereas the correlation between the $\gamma$-ray and X-ray bands appears more complex in nature.
This complexity can be qualitatively connected to the lower fractional variability observed in the X-ray band (subsection\,\ref{sec:3.5}). Since the X-ray emission is likely dominated by low-energy electrons, its response to variations in the seed photon fields is expected to be weaker and slower than that of the $\gamma$-ray emission, which is governed by IC scattering involving higher-energy electrons. As a result, the different variability amplitudes and response timescales in the two bands may lead to multi-lag correlation rather than a single, well-defined correlation delay.

The temporal analysis of the multi-wavelength lightcurves indicates that the source exhibits significant variability across all energy bands, with the lowest variability observed in the X-ray band and the highest variability in the $\gamma$-ray band (see, Fig.\,\ref{fig: fvar}). The reduced variability in the X-ray band and the enhanced variability in the UV/optical and $\gamma$-ray bands can be naturally interpreted in the context of the energy distribution of relativistic electrons and the characteristic double-hump structure of blazar spectral energy distributions. The UV/optical and $\gamma$-ray bands are located near the peaks of the broadband SED, where the emission is dominated by high-energy electrons through synchrotron and inverse Compton processes (with both IR and BLR seed photons contributing in this work), respectively. In contrast, the X-ray emission at keV energies typically lies on the rising part or tail of the inverse Compton component and is mainly produced via synchrotron self-Compton scattering by relatively low-energy electrons. Consequently, the UV/optical and GeV emissions are governed by higher-energy electrons, while the X-ray emission is dominated by lower-energy electron population, leading to a comparatively reduced variability amplitude in the X-ray regime. The observed highest variability amplitude in the $\gamma$-ray band relative to the X-ray and optical/UV bands is consistent with results from studies of other blazars \citep[e.g.,][]{2005ApJ...629..686Z,2019MNRAS.484.3168S}.  
This behavior can also be attributed to the more rapid radiative cooling of high-energy electrons compared to their low-energy counterparts, a scenario that is consistent with the symmetric  flare profiles observed in the $\gamma$-ray band (see, subsection\,\ref{sec:3.1}). Moreover, the increase in variability amplitude with photon energy may also reflect intrinsic spectral variability within the source \citep{2005ApJ...629..686Z}.


To further investigate the underlying mechanism behind the $\gamma$-ray flares of S5\,1044+71, we model the simultaneous multiband SEDs across the different flux states. The broadband SED  in different flux states is well reproduced by a standard one-zone leptonic model including synchrotron, SSC and EC components. In our fits, the high-energy $\gamma$-ray emission cannot be accounted for by either EC scattering of IR photons or EC scattering of BLR photons. Instead, both IR and BLR seed photon fields are required to reproduce the observed emission. In particular, we find that EC scattering of infrared  photons extends upto the very-high-energy (VHE) end of the SED (avoiding Klein–Nishina suppression), while EC scattering of BLR photons sufferes KN suppression and accounts mostly to the GeV flux.
This two-component EC scenario is  invoked in other powerful FSRQs to match their double-humped SEDs (see, e.g., \cite{2014ApJ...790..147B}).
If we assume $\delta=20$, the emission region can be located at $d_{\gamma}\simeq 0.0292$pc from the nucleus. This is comparable to the canonical BLR radius ($\sim 3.4 \times 10^{17}\,\mathrm{cm}$) \citep{2024ApJ...962...22Z}. This suggests that the emission region is likely located near the BLR, where both BLR and IR photon fields are sufficiently dense to contribute to the EC scattering process.
We observe systematic trends in the model parameters from low to high flux states. In brighter states the particle energy distribution becomes flatter (indicating a harder spectrum), the electron break energy $\gamma_b$ shifts to larger values, and the magnetic field strength B is reduced. Physically, a harder electron index and higher $\gamma_b$ imply that the jet’s acceleration processes are injecting a larger fraction of high-energy electrons during flares. At the same time, a weaker B-field reduces synchrotron cooling and makes the jet more particle-dominated. This combination naturally pushes both SED peaks to higher frequencies and increases the Compton dominance. Qualitatively, our result  a ``harder‐when‐brighter” electron spectrum coupled with lower B in the high‐flux states – is consistent with similar findings in other blazars. For example, multi‐epoch SED modeling of 3C\,279 showed that the high state exhibited a slightly reduced magnetic field and a flatter electron spectrum compared to the quiescent state \citep{2019MNRAS.484.3168S,2022PASP..134j4101W,2022APh...13902687T}.  Such behavior has also been seen in PKS\,1510-089 and CTA\,102 flares, where flaring states required an injection of very energetic electrons (high $\gamma_b$) and sometimes a lower B. In contrast, some other studies (e.g., \cite{2025arXiv250418927M} on FSRQ 4C+27.50) report increases in B and Doppler factor during flares, underscoring that multiple mechanisms may drive different outbursts. In our case, the consistently higher $\gamma_b$ and harder spectra at high flux imply more efficient acceleration when the jet is most active.  Since we infer the emission region lies near the BLR, the reduced B could also indicate the flaring zone moves slightly beyond the inner BLR, where field strength drops; in that case IC scattering of IR photons could further dominate. 
The dominance of IC scattering of IR photons is further supported by the detection of $\sim$40–50 GeV photons (see subsection\,\ref{sec:3.2}), which escape without significant $\gamma$-$\gamma$ absorption, also suggesting that the emission region is not located deep within the BLR, but is instead situated at or beyond BLR.
Overall, the parameter evolution – harder electron spectrum, weaker B and higher break energy in the high state – is qualitatively consistent with a picture where during outbursts a larger fraction of jet power is channeled into relativistic electrons. Future observations of the source, particularly in the  VHE $\gamma$-ray regime, will be crucial for rigorously testing these scenarios, as they will probably better constrain the location of the emission region and the dominant radiative processes during different activity states.


\section*{Acknowledgements}

The authors thank the anonymous referee for valuable comments and suggestions. The authors thank the Department of Physics, University of Kashmir for providing the necessary facilities to carry out this research work. ZS is supported by the Department of Science and Technology (DST), Govt. of India, under the INSPIRE Faculty grant (DST/INSPIRE/04/2020/002319). ZM acknowledges the financial support provided by the Science and Engineering Research Board (SERB), Government of India, under the National Postdoctoral Fellowship (NPDF), Fellowship reference no. PDF/2023/002995. SA also thanks IUCAA for providing the facilities. This research has made use of $\gamma$-ray data from \emph{Fermi} Science Support Center (FSSC). The work has also used the Swift Data from the High Energy Astrophysics Science Archive Research Center (HEASARC), at NASA’s Goddard Space Flight Center.

\section*{Data Availability}

The data and the model used in this article will be shared on
reasonable request to the corresponding author, Sajad Ahanger (email: sajadphysics21@gmail.com)





%

\bibliography{sample701}{}
\bibliographystyle{aasjournalv7}



\end{document}